\documentclass[12pt]{article}
\usepackage{rotating} 
\usepackage{epstopdf}
\usepackage{adjustbox,lipsum}
\usepackage{float}
\usepackage{amssymb}
\usepackage[english]{babel}
\usepackage{amsthm}
\usepackage[table]{xcolor}
\usepackage{booktabs}
\usepackage{amsmath}
\usepackage{latexsym}
\usepackage[noend]{algpseudocode}
\usepackage{epsfig}
\usepackage{graphicx}
\usepackage{wasysym}
\usepackage{threeparttable}
\usepackage{natbib}
\usepackage{color}
\usepackage{algorithm}
\usepackage[noend]{algpseudocode}
\usepackage{epstopdf}
\usepackage{caption}
\usepackage{subcaption}
\usepackage{booktabs}

\usepackage{multirow}
\usepackage[breaklinks]{hyperref}

\usepackage{enumitem}
\usepackage{caption}
\usepackage{algorithm}
\def\Tiny{\fontsize{10pt}{10pt}\selectfont}
\def\boxit#1{\vbox{\hrule\hbox{\vrule\kern6pt
          \vbox{\kern6pt#1\kern6pt}\kern6pt\vrule}\hrule}}

\def\bse{\begin{eqnarray*}}
\def\ese{\end{eqnarray*}}
\def\be{\begin{eqnarray}}
\def\ee{\end{eqnarray}}
\def\bq{\begin{equation}}
\def\eq{\end{equation}}
\def\bse{\begin{eqnarray*}}
\def\ese{\end{eqnarray*}}

\usepackage{mathptmx}      
\usepackage{bm}

 {\begin{list}{}%
         {\setlength{\leftmargin}{#1}}%
         \item[]%
 }
 {\end{list}}
\usepackage{setspace}

\newtheorem{theorem}{Theorem}[section]

\begin{document}
\thispagestyle{empty} \baselineskip=28pt

\begin{center}
{\LARGE{\bf Generating Independent Replicates Directly from the Posterior Distribution for a Class of Spatial Latent Gaussian Process Models}}
\end{center}

\baselineskip=12pt

\vskip 2mm
\begin{center}
 Jonathan R. Bradley\footnote{(\baselineskip=10pt to whom correspondence should be addressed) Department of Statistics, Florida State University, 117 N. Woodward Ave., Tallahassee, FL 32306-4330, jrbradley@fsu.edu} and 
 Madelyn Clinch\footnote{Department of Statistics, Florida State University, 117 N. Woodward Ave., Tallahassee, FL 32306-4330}
\end{center}
%
%
%
%
\vskip 4mm

\begin{center}
\large{{\bf Abstract}}
\end{center}
Markov chain Monte Carlo (MCMC) allows one to generate dependent replicates from a posterior distribution for effectively any Bayesian hierarchical model. However, MCMC can produce a significant computational burden. This motivates us to consider finding expressions of the posterior distribution that are computationally straightforward to obtain independent replicates from directly. We focus on a broad class of Bayesian latent Gaussian process (LGP) models that allow for spatially dependent data. First, we derive a new class of distributions we refer to as the generalized conjugate multivariate (GCM) distribution. The GCM distribution's theoretical development is similar to that of the CM distribution with two main differences; namely, (1) the GCM allows for latent Gaussian process assumptions, and (2) the GCM explicitly accounts for hyperparameters through marginalization. The development of GCM is needed to obtain independent replicates directly from the exact posterior distribution, which has an efficient projection/regression form. Hence, we refer to our method as Exact Posterior Regression (EPR). Illustrative examples are provided including simulation studies for weakly stationary spatial processes and spatial basis function expansions. An additional analysis of poverty incidence data from the U.S. Census Bureau's American Community Survey (ACS) using a conditional autoregressive model is presented.
\baselineskip=12pt

%

\baselineskip=12pt
\par\vfill\noindent
\par\medskip\noindent
\pagebreak\newpage \pagenumbering{arabic}
\baselineskip=24pt
\singlespacing
\section{Introduction} 

MCMC has become an invaluable tool in statistics and is covered in standard text books \citep{mcmc-book-casella}. MCMC is an all purpose strategy that allows one to obtain dependent samples from a generic posterior distribution. There are several theoretical considerations that one needs to consider when implementing MCMC to obtain samples from the posterior distribution including, ergodicity, irreducibility, and positive recurrence of the MCMC. In addition to theoretical considerations, practical implementation issues arise, including, a potential for high computational costs, assessing convergence \citep{gr1,cowles1996markov}, tuning the MCMC \citep{roberts2009examples}, and computing the effective sample size of the Markov chain \citep{vats2019multivariate}, among other considerations. One of the current state-of-the-art techniques in MCMC is Hamiltonian Monte Carlo \citep[HMC,][]{neal2011mcmc}. HMC is a Metropolis$-$Hastings algorithm, where Hamiltonian dynamic evolution is used to propose a new value. In general, HMC leads to ``fast mixing'' (i.e., converges relatively quickly to the posterior distribution) because it provides a sample from the joint posterior distribution of all processes and parameters, and moreover, has been optimized efficiently using the software \texttt{Stan} \citep{carpenter2017stan}.

Of course, MCMC is not needed if one can obtain independent replicates directly from the posterior distribution efficiently. In this article, we revisit the problem of generating independent replicates directly from the posterior distribution for a broad class of spatial latent Gaussian process models \citep[LGP, e.g., see][]{gs}. Much of the current literature does not consider solving this problem, since obtaining independent replicates directly from the exact posterior distribution for Bayesian spatial LGPs is a difficult problem, and MCMC can easily be adapted to many settings.  We consider Bayesian spatial LGPs for Gaussian distributed data, Poisson distributed data, and binomial distributed data. The samples from our proposed model are independently drawn, and hence avoid issues with convergence, tuning, and positive autocorrelations in a MCMC. Moreover, our exact replicates have an interpretable projection formulation. This regression-type projection can be computed efficiently using known block matrix inversion formulas \citep{lu2002inverses}. Thus, we refer to our method as Exact Posterior Regression (EPR), which is the one of the contributions of this article.

Conjugate prior distributions are often restricted to the data type. For example, for binomial, negative binomial, Bernoulli, and multinomial distributed data, the fixed and random effects are conjugate with the multivariate logit-beta distribution \citep{gao2019bayesian,bradley2019spatio}, which is the special case of the conjugate multivariate (CM) distribution. Similarly, Poisson and Weibull distributed data are conjugate with the multivariate log-gamma distribution \citep{bradleyPMSTM,hu2018bayesian,xu2019latent,yang2019bayesian,parker2020conjugate,parker2021general}, another special case of the CM distribution. Finally, mixed effects models for Gaussian distributed data regularly make use of Gaussian priors for fixed and random effects \citep{gelmanbook}, which is also a type of CM distribution. Thus, our second major contribution is to extend the conjugate multivariate (CM) distribution \citep{bradleyLCM} to LGPs. Additionally, conjugate prior distributions and the CM distribution do not allow one to explicitly account for hyperparameters without the use of MCMC or approximate Bayesian techniques. Thus, in our extension of the CM to LGPs we marginalize across hyperparameters. We call this new distribution the generalized CM (GCM) distribution, which allows for standard latent Gaussian process model specifications of spatial LGPs \citep[e.g., see][for a recent discussion]{gs}. Furthermore, we develop conditional distributions for GCM distributed random vectors. 

A key step in our formulation is the incorporation of what we call ``discrepancy term,'' which are simply additive term introduced into a mixed effects model similar to that of \citet{bradleyhierarchical} and \citet{bradley2023deep}. Classical spatial LGP models set these parameters equal to zero. When these terms are not set equal to zero and instead given a type of improper prior then we show that the implied posterior distribution for fixed and random effects is of the form of a GCM, which we can directly sample from (bypassing the need for MCMC). However, we show that posterior replicates from this GCM can overfit the latent process. Thus, we suggest including these discrepancy parameters in the model to bypass MCMC, and then marginalize them from the posterior distribution and estimate them to be zero.
%

We emphasize the high potential impact of the contributions of EPR and GCM, since much of the literature places a high consistent emphasis on using MCMC strategies to obtain asymptotically exact correlated samples from the posterior distribution. For example, at the time of writing this manuscript the following papers use MCMC in a spatial LGP setting: \citet{kang2023fast}, \citet{konomi2023bayesian}, \citet{porter2023objective}, \citet{vranckx2023spatial}, and \citet{zhang2023bayesian}, among others. All of these analyses can easily be adapted to be implemented using EPR, which completely avoids MCMC.

EPR allows one to efficiently analyze several types of correlated spatial data. In particular, we consider  modeling three ``types of data,'' namely,  conditionally Gaussian, Poisson, and binomial distributed spatial data. Computationally expensive MCMC techniques have become a standard for modeling spatial data \citep{robert2011short, gs}. Also, a common approximate Bayesian technique used frequently in the spatial statistics literature is referred to as integrated nested Laplace approximations \citep[INLA,][]{lindgren2022spde}. In this article, we compare MCMC and INLA to EPR.  

To summarize, the contributions of this article can be classified into two groups:
\begin{enumerate}
	\item The first group of contributions of this article develops the GCM distribution. This includes integral expressions for the GCM distribution and the conditional GCM distribution up to a proportionality constant. The key literature on conjugate modeling began with \citet{diaconis}'s seminal paper which formally developed univariate conjugate models for the exponential family. Then \citet{chen2003conjugate} developed \citet{diaconis}'s work in the context of fixed effects models and \citet{bradleyLCM} developed \citet{diaconis}'s work in the context of mixed effects models. However, all of these papers require one to match the form of the prior distribution with that of the likelihood. The use of the GCM allows one to consider LGPs. Moreover, this literature often does not emphasize hyperparameters; however, our development explicitly addresses hyperparameters through marginalization. It should be noted that the theoretical development of the GCM is similar to that of the CM distribution \citep{bradleyLCM}. However, the GCM has an enormous practical advantage over the CM by allowing one to use a more standard class (i.e., LGP) of process and prior distributions for spatial data and avoids MCMC updates of hyperparameters. For example, when using the CM for a Poisson data settings, one uses multivariate log-gamma priors for fixed and random effects and updates shape/rate parameters in an MCMC. When using the GCM one can use Gaussian priors and avoid sampling hyperparameters in an MCMC.
	\item The second group of contributions of this article allows one to use the GCM in a Bayesian LGP context to produce what we call exact posterior regression (EPR). Much of the Bayesian literature is shifting its' focus on avoiding MCMC through the use of approximate Bayesian methods \citep[e.g., see][]{wainwright2008graphical,rue} or through direct sampling of the posterior distributions in special cases for Gaussian data \citep{zhang2021high,van2021fast,shirota2023conjugate,zhang2023exact}. Recently, \citet{bradley2023deep} developed an exact sampler from the posterior distribution for a deep Bayesian statistical model for Gaussian and non-Gaussian spatio-temporal data referred to as the deep hierarchical generalized transformation model. EPR adds to this growing literature by allowing one to independently sample from the posterior from a broad class of spatial LGPs. By ``broad'' we mean that many existing spatial LGPs can be written in terms of our formulation. We show that the posterior distribution for fixed and random effects are GCM. Furthermore, we use matrix algebra techniques to aid in the computation of EPR (see Theorems~\ref{thm:5} and \ref{thm:7}).
\end{enumerate}
\noindent
The remainder of the article proceeds as follows. Before we introduce our proposed LGP, we will first provide derivations of the GCM and conditional GCM distribution in Section~\ref{sec:gcm}. We emphasize that that GCM random vectors are derived through how they are simulated. Then, in Section~\ref{sec:epr} we show that our proposed model's posterior distribution is GCM, and we describe how to efficiently sample independent replicates directly from the marginal posterior of the fixed effects, and random effects (which we call EPR). Illustrations are provided in Section~\ref{sec:empr}, which includes several simulations/comparisons (15 in total) including common models used in spatial statistics: weakly stationary spatial processes, spatial basis function expansions, and conditional autoregressive models. The main goal of our illustrations is to compare to several common existing strategies for Bayesian spatial LGPs. Proofs are given in the Appendix, and a discussion is given in Section~\ref{sec:discussion}.


\section{Preliminary Derivations: The Generalized Conjugate Multivariate Distribution}\label{sec:gcm}

We now derive the \textit{generalized conjugate multivariate} (GCM) distribution. This development is  similar to the development of the CM distribution from \citet{bradleyLCM}. We give a review of the CM distribution in Appendix A. The difference between the GCM and CM is that the GCM drops the assumption of identical classes of DY random variables, and marginalizes across a generic $d$-dimensional real-valued parameter vector $\bm{\theta}$. The GCM is needed for our main contribution of EPR in Section 4. 

The GCM is defined by the transformation,
\begin{equation}\label{cholV2}
\textbf{y} = \bm{\mu}_{M} + \textbf{V}_{M}\textbf{D}(\bm{\theta})\textbf{w}_{M},
\end{equation}
\noindent
where the $n\equiv\sum_{k = 1}^{K}n_{k}$-dimensional random vector $\textbf{y} = (\textbf{y}_{1}^{\prime},\ldots,\textbf{y}_{K}^{\prime})^{\prime}$, $n_{k}$-dimensional random vector $\textbf{y}_{k} = (Y_{k,1},\ldots, Y_{k,n_{k}})^{\prime}$, the $n$-dimensional random vector $\textbf{w}_{M} = (\textbf{w}_{1}^{\prime},\ldots, \textbf{w}_{K}^{\prime})^{\prime}$ with $(k,i)$-th element $w_{k,i} \sim \mathrm{DY}(\alpha_{k,i},\kappa_{k,i};\psi_{k})$, ``DY'' is a shorthand for the well known Diaconis-Ylvisaker random variable \citep{diaconis} (see Appendix A for a review), the subscript ``M'' stands for ``Multi-type,'' the $n\times n$ real-valued matrix $\textbf{V}_{M}$ is an invertable covariance parameter matrix, and $\bm{\mu}_{M}$ is an unknown $n$-dimensional real-valued location parameter vector. The function $\psi_{k}$ is referred to as the unit log partition function, and we consider $\psi_{1}(w) = {w}^{2}$, $\psi_{2}(w) = \mathrm{exp}(w)$, and $\psi_{3}(w) = \mathrm{log}\{1+\mathrm{exp}(w)\}$ for real-valued $w$. It is known that $w_{1,i}$ is normally distributed with mean $\frac{\alpha_{1,i}}{2\kappa_{1,i}}$ and variance $\frac{1}{2\kappa_{1,i}}$, $w_{2,i}$ is the log of a gamma random variable with shape $\alpha_{2,i}$ and rate $\kappa_{2,i}$, and $w_{3,i}$ is the logit of a beta random variable with shape parameters $\alpha_{3,i}$ and rate $\kappa_{3,i}-\alpha_{3,i}$ \citep{bradleyLCM}.

 Let $\textbf{D}: \Omega \rightarrow \mathbb{R}^{n}\times \mathbb{R}^{n}$ be a known $n\times n$ matrix valued function, such that $\textbf{D}(\bm{\theta})^{-1}$ exists for every $d$-dimensional $\bm{\theta}\in \Omega$ for a generic real-valued set $\Omega$. Let $\bm{\theta}$ be distributed according to the proper density $\pi(\bm{\theta})$, where $\bm{\theta}$ is independent of $\bm{\mu}_{M}$, $\bm{\alpha}_{M}$, $\bm{\kappa}_{M}$, and $\textbf{V}_{M}$. Sampling from the marginal distribution $\textbf{y}\vert \bm{\mu}_{M}, \textbf{V}_{M},\bm{\alpha}_{M},\bm{\kappa}_{M}$  (marginalizing across $\bm{\theta}$) is straightforward; namely, first sample $\bm{\theta}$ from $\pi(\bm{\theta})$ and then compute the transformation in (\ref{cholV2}) to produce a sample from $f(\textbf{y}\vert \bm{\mu}_{M},\textbf{V}_{M}, \bm{\alpha}_{M},\bm{\kappa}_{M})$. The pdf $\textbf{y}\vert \bm{\mu}_{M},\textbf{V}_{M}, \bm{\alpha}_{M},\bm{\kappa}_{M}$ is stated in Theorem~\ref{thm:1}.\\

\begin{theorem}\label{thm:1}
Let $\textbf{y}$ be defined as in (\ref{cholV2}). Then the pdf for $\textbf{y}$ is given by,
\begin{align}
\label{mlg_pdf2}
\nonumber
& f(\textbf{y}\vert \bm{\mu}_{M},\textbf{V}_{M},\bm{\alpha}_{M},\bm{\kappa}_{M}) \\ &=\int_{\Omega}\pi(\bm{\theta})\mathcal{N}_{M}\hspace{2pt}\mathrm{exp}\left[\bm{\alpha}_{M}^{\prime}\textbf{D}(\bm{\theta})^{-1}\textbf{V}_{M}^{-1}(\textbf{y} - \bm{\mu}_{M}) - \bm{\kappa}_{M}^{\prime}\bm{\psi}_{M}\left\lbrace\textbf{D}(\bm{\theta})^{-1}\textbf{V}_{M}^{-1}(\textbf{y} - \bm{\mu}_{M})\right\rbrace\right]d\bm{\theta},
\end{align}
where $\mathcal{N}_{M} = \frac{\left\lbrace\prod_{k = 1}^{K}\prod_{i = 1}^{n_{k}}\mathcal{N}_{k}(\kappa_{k,i},{\alpha_{k,i}})\right\rbrace}{\mathrm{det}\left\lbrace\textbf{D}(\bm{\theta})\right\rbrace\mathrm{det}(\textbf{V}_{M})}$, $\textbf{y}\in \mathcal{S}$, $\mathcal{S} = \{\textbf{y}: \textbf{y} = \bm{\mu}_{M} + \textbf{V}_{M}\textbf{D}(\bm{\theta})\textbf{c}, \textbf{c} = \{c_{k,i}\}, c_{k,i}\in \mathcal{Y}_{k}, \bm{\theta}\in \Omega,i = 1,\ldots, n_{k},k = 1,\ldots, K\}$, $\alpha_{k,i}/\kappa_{k,i} \in \mathcal{Z}_{k}$, $\kappa_{k,i}>0$, $\bm{\psi}_{M}\left\lbrace\textbf{V}_{M}(\textbf{y} - \bm{\mu}_{M})\right\rbrace = \left(\psi_{1}\left\lbrace\textbf{J}_{1}\textbf{V}_{M}(\textbf{y} - \bm{\mu}_{M})\right\rbrace^{\prime},\right.$\\ $\left.\ldots, \psi_{K}\left\lbrace\textbf{J}_{K}\textbf{V}_{M}(\textbf{y} - \bm{\mu}_{M})\right\rbrace^{\prime}\right)^{\prime}$, the $n_{k}\times n$ matrix $\textbf{J}_{k} = \left(\bm{0}_{n_{k},\sum_{j = 1}^{k-1}n_{j}},\textbf{I}_{n_{k}}, \bm{0}_{n_{k},\sum_{j = k+1}^{K}n_{j}}\right)$, $\bm{0}_{n,m}$ is an $n\times m$ matrix of zeros, $\textbf{I}_{n_{k}}$ is an $n_{k}\times n_{k}$ identity matrix, the $n$-dimensional vector $\bm{\alpha}_{M} = (\bm{\alpha}_{1}^{\prime},\ldots, \bm{\alpha}_{K}^{\prime})^{\prime}$, and the $n$-dimensional vector $\bm{\kappa}_{M} = (\bm{\kappa}_{1}^{\prime},\ldots, \bm{\kappa}_{K}^{\prime})^{\prime}$. 
\end{theorem}

\noindent
\textit{Proof:} See Appendix B.\\

\noindent
 We use the shorthand $\mathrm{GCM}(\bm{\alpha}_{M},\bm{\kappa}_{M},\bm{\mu}_{M},\textbf{V}_{M},\pi,\textbf{D}; \bm{\psi}_{M})$ for the density in (\ref{mlg_pdf2}).  
 
 Sampling directly from a GCM distribution requires two items:
 \begin{enumerate}
 	\item One must be able to sample the random vector $\bm{\theta}$ directly from its prior distribution $\pi$.
 	\item One must be able to the sample independent DY random variables contained in the vector $\textbf{w}_{M}$.
 \end{enumerate}
In this article, the parameter vector $\bm{\theta}$ typically consists of variance parameters and spatial range parameters. These parameters will be given independent inverse gamma prior or uniform prior distributions, which one can sample from directly. Additionally, the class of LGPs in Section~\ref{sec:epr} lead to DY random variables that are either independent univariate normal, beta, or gamma random variables, which are straightforward to simulate from directly using standard software. The fact that we can sample independent replicates of a GCM random vector directly is crucial in Section~\ref{sec:epr}, where we show that a certain class of LGPs leads to a posterior distribution that is GCM (i.e., is of the form in Theorem~\ref{thm:1}), and hence, one can directly sample from it.
 
%
 
 We now provide the integral expression for the conditional GCM in Theorem~\ref{thm:2} up to a proportionality constant.
\begin{theorem}\label{thm:2}
{Let $\textbf{y} = (\textbf{y}^{(1)\prime},\textbf{y}^{(2)\prime})^{\prime}\sim \mathrm{GCM}(\bm{\alpha}_{M},\bm{\kappa}_{M},\bm{\mu}_{M},\textbf{V}_{M},\pi,\textbf{D}; \bm{\psi}_{M})$, where $\textbf{y}^{(1)}$ is $r$-dimensional and $\textbf{y}^{(2)}$ is $(n-r)$-dimensional. Also, let $\textbf{V}_{M}^{-1} = (\textbf{H}, \textbf{Q})$, where $\textbf{H}$ is a $n\times r$ and $\textbf{Q}$ is $n\times (n-r)$. Then, it follows}
\begin{align}
\nonumber
& f(\textbf{y}^{(1)}\vert \textbf{y}^{(2)}, \bm{\mu}_{M},\textbf{V}_{M},\bm{\alpha}_{M},\bm{\kappa}_{M})\\
\nonumber
& \propto \int_{\Omega} \frac{\pi(\bm{\theta})}{\mathrm{det}\left\lbrace \textbf{D}(\bm{\theta})\right\rbrace}\mathrm{exp}\left[\bm{\alpha}_{M}^{\prime}\textbf{D}(\bm{\theta})^{-1}\textbf{H}\textbf{y}^{(1)}-\bm{\alpha}_{M}^{\prime}\bm{\mu}_{M}^{*} - \bm{\kappa}_{M}^{\prime}\bm{\psi}_{M}\left\lbrace\textbf{D}(\bm{\theta})^{-1}\textbf{H}\textbf{y}^{(1)} - \bm{\mu}_{M}^{*}\right\rbrace\right]d\bm{\theta},
\end{align}
\noindent
{where $\bm{\mu}_{M}^{*} = \textbf{D}(\bm{\theta})^{-1}\textbf{V}_{M}^{-1}\bm{\mu}_{M}-\textbf{D}(\bm{\theta})^{-1}\textbf{Q}\textbf{y}^{(2)}$.}
\end{theorem}
\noindent
\textit{Proof:} See Appendix B.\\

\noindent
We use the shorthand cGCM$(\bm{\alpha}_{M},\bm{\kappa}_{M},\bm{\mu}_{M}^{*}, \textbf{H}, \pi, \textbf{D};\bm{\psi}_{M})$ for the conditional GCM in Theorem~\ref{thm:2}. It is not known how to simulate directly from a cGCM.\\

\section{Methodology}\label{sec:epr}

In this section, we outline how to sample from the posterior distribution of fixed and random effects from a general class of spatial LGPs. We define EPR in Section~\ref{sec:epr:nongauss} for areal spatial data, define the extension to spatial process models in Section~\ref{sec:process}, and discuss computational issues in Section~\ref{sec:comp}.

\subsection{Exact Posterior Regression for Regional Data}\label{sec:epr:nongauss}

Suppose we observe data from the exponential family, let the total number of observations be denoted with $n$, and denote the $n$-dimensional data vector with $\textbf{z} = (Z_{1},\ldots,Z_{n})^{\prime}$. Let $Z_{i}$ represent the data at region $i$ (e.g., counties, census tracts, etc.). Then assume $Z_{i}$ belongs to a member of the exponential family of distributions. In particular, we assume {one} of the following: 
\begin{align}\label{EF2}
	Z_{i}\vert Y_{i},b_{i,k} & \sim \mathrm{EF}(Y_{i}, b_{i,k}, \psi_{k}); \hspace{2pt} i = 1,\ldots, n, \hspace{2pt} k = 1,2,3
\end{align}
\noindent
where ``EF'' is a shorthand for the natural exponential family (see Appendix A for more details), and $b_{i,k}\psi_{k}(Y_{i})$ is the log-partition function. For example, when $b_{i,1} = \frac{1}{2\sigma_{i}^{2}}$ with $\sigma_{i}^{2}>0$ and $\psi_{1}(Y_{i}) = Y_{i}$ we have that $Z_{i}\vert Y_{i},b_{i,1}$ is normally distributed with mean $Y_{i}$ and variance $\sigma_{i}^{2}$. When $b_{i,2} \equiv 1$ and $\psi_{2}(Y_{i}) = \mathrm{exp}(Y_{i})$ we have that $Z_{i}\vert Y_{i},b_{i,2}$ is Poisson distributed with mean $\mathrm{exp}(Y_{i})$. Similarly, when $b_{i,3} = m_{i}$ with integer $m_{i}\ge 1$ and $\psi_{3}(Y_{i}) = \mathrm{log}\{1+\mathrm{exp}(Y_{i})\}$ we have that $Z_{i}\vert Y_{i},b_{i,3}$ is binomial distributed with sample size $m_{i}$ and probability of success $\mathrm{exp}(Y_{i})/\{1+\mathrm{exp}(Y_{i})\}$. In this article, we consider these three cases (i.e., normal, Poisson and binomial distributed cases), and note that binomial distributed data allows for Bernoulli distributed data as a special case (i.e., $m_{i}= 1$), and multinomial distributed data when using a stick-breaking representation of the multinomial \citep[e.g., see][for stick-breaking in the context of CM prior distributions]{bradley2019spatio}. 

Now, organize the latent random variable $Y_{i}$ into the $n$-dimensional vector $\textbf{y} = (Y_{1},\ldots, Y_{n})^{\prime}$. Consider the following linear model assumption for $\textbf{y}$ \citep{glm-nelder}:
\begin{align}\label{mem2}
\textbf{y} = \textbf{X}\bm{\beta} + \textbf{G}\bm{\eta} + (\bm{\xi}-\bm{\delta}_{y}),
\end{align}
\noindent
where $\textbf{X}$ is a $n\times p$ matrix of known covariates, and $\bm{\beta}$ is an unknown $p$-dimensional vector of regression coefficients. Let $\bm{\beta}$ have a Gaussian prior with $p$-dimensional location vector $\bm{\delta}_{\beta}$, and $p\times p$ covariance matrix $\textbf{D}_{\beta}(\bm{\theta})\textbf{D}_{\beta}(\bm{\theta})^{\prime}$, where $\textbf{D}_{\beta}(\bm{\theta}): \Omega \rightarrow \mathbb{R}^{p}\times \mathbb{R}^{p}$. Let $\textbf{G}$ be a $n\times r$ matrix of coefficients for the $r$-dimensional random effects $\bm{\eta}$. In this article, $\textbf{G}$ will be set equal to a known pre-specified matrix of basis functions (e.g., splines \citep{wahba}, wavelets \citep{wavelet}, Moran's I basis functions \citep{hughes}, etc.), or a matrix square root of a known spatial covariance matrix. We assume $\bm{\eta}$ is Gaussian with $r$-dimensional location vector $\bm{\delta}_{\eta}$ and $r\times r$ covariance matrix $\textbf{D}_{\eta}(\bm{\theta})\textbf{D}_{\eta}(\bm{\theta})^{\prime}$, where $ \textbf{D}_{\eta}(\bm{\theta}): \Omega \rightarrow \mathbb{R}^{r}\times \mathbb{R}^{r}$. Let $\bm{\theta}$ be a generic $d$-dimensional parameter vector with prior distribution $\pi(\bm{\theta})$.

Traditionally, the fine-scale variability term $\bm{\xi}$ is assumed to be Gaussian \citep{cressie-wikle-book}. In our framework, it will prove to be useful to specify the distribution for $\bm{\xi}$ to be a cGCM that is ``close'' to a Gaussian distribution. Specifically, let the distribution for $\bm{\xi}$ be proportional to a cGCM$(\bm{\alpha}_{\xi},\bm{\kappa}_{\xi},\bm{\delta}_{\xi}^{*}, \textbf{H}_{\xi}, \pi_{\xi}, \textbf{D}_{\xi};\bm{\psi}_{\xi})$, where the $2n$-dimensional discrepancy parameter $\bm{\delta}_{\xi}^{*}=(\bm{\delta}_{y}^{\prime}-\bm{\beta}^{\prime}\textbf{X}^{\prime} - \bm{\eta}^{\prime}\textbf{G}^{\prime},\bm{\delta}_{\xi}^{\prime})^{\prime}$, $\bm{\delta}_{y}$ and $\bm{\delta}_{\xi}$ are $n$-dimensional real-vectors, and $2n\times n$ matrix-valued precision parameter $\textbf{H}_{\xi}=(\textbf{I}_{n},\sigma_{\xi}^{2}\textbf{I}_{n})^{\prime}$. The $2n$-dimensional shape parameter $\bm{\alpha}_{\xi} = \bm{0}_{2n,1}$ when the data is assumed Gaussian, and $\bm{\alpha}_{\xi} = (\alpha_{\xi}\bm{1}_{1,n},\bm{0}_{1,n})^{\prime}$ when the data is assumed to be distributed according to the Poisson or binomial distributions, where $\alpha_{\xi}>0$ and $\bm{1}_{r,n}$ is a $r\times n$ matrix of ones. The $2n$-dimensional shape parameter $\bm{\kappa}_{\xi} = (\bm{0}_{1,n}, \frac{1}{2}\bm{1}_{1,n})^{\prime}$ when the data is assumed to be either Gaussian or Poisson distributed, and $\bm{\kappa}_{\xi} =(2\alpha_{\xi}\bm{1}_{1,n}, \frac{1}{2}\bm{1}_{1,n})^{\prime}$ when the data is assumed to be distributed according to the binomial distribution. Let $\textbf{D}_{\xi}\equiv \sigma_{\xi}^{2}\textbf{I}_{2n}$ with $\sigma_{\xi}^{2}>0$. The unit-log partition function $\bm{\psi}_{\xi}$ is, 
\begin{equation}
\nonumber
\bm{\psi}_{\xi}(\textbf{h}) = (\psi_{k}(h_{1}),\ldots,\psi_{k}(h_{n}),\psi_{1}(h_{1}^{*}),\ldots, \psi_{1}(h_{n}^{*}))^{\prime},
\end{equation}
for any $\textbf{h} = (h_{1},\ldots, h_{n},h_{1}^{*},\ldots, h_{n}^{*})^{\prime}\in \mathbb{R}^{2n}$. It is straightforward to verify that when $\alpha_{\xi}=0$ we have that cGCM$(\bm{\alpha}_{\xi},\bm{\kappa}_{\xi},\bm{\delta}_{\xi}^{*}, \textbf{H}_{\xi}, \pi_{\xi}, \textbf{D}_{\xi};\bm{\psi}_{\xi})$ is proportional to a Gaussian distribution with mean $\bm{\delta}_{\xi}$ and covariance $\sigma_{\xi}^{2}\textbf{I}_{n}$ with $\sigma_{\xi} \in \bm{\theta}$. This choice of cGCM with $\alpha_{\xi}>0$ will ensure that the implied posterior distribution has parameters that do not lie on the boundary of the parameter space.

The terms $\textbf{X}\bm{\beta}$, $\textbf{G}\bm{\eta}$, and $\bm{\xi}$ are covered in standard textbooks in spatio-temporal statistics \citep{cressie-wikle-book}, and are referred to as large-scale variability, small-scale variability, and fine-scale variability, respectively. In more recent literature a fourth term has been considered \citep{bradleyhierarchical,bradley2023deep};  that is, the $(2n+p+r)$-dimensional vector $\bm{\delta} = (\bm{\delta}_{y}^{\prime},\bm{\delta}_{\beta}^{\prime},\bm{\delta}_{\eta}^{\prime},$ $\bm{\delta}_{\xi}^{\prime})^{\prime}$ discrepancy parameter. These discrepancy parameters often lead to more efficient procedures to sample from the posterior distribution. In our case, a \textit{particular form} of $\bm{\delta}$ leads the fixed and random effects to be distributed according to a GCM, which from Section~\ref{sec:gcm}, we know how to sample from directly without approximations and without MCMC. Specifically, let $\bm{\delta} = -\textbf{D}(\bm{\theta})^{-1}\textbf{Q}\textbf{q}$, where $\textbf{Q}$ are the $(2n+p+r)\times n$ eigenvectors of the orthogonal complement of the $(2n+p+r)\times (n+p+r)$ matrix,
\begin{equation}\label{standardH}
\textbf{H} = \left(\begin{array}{ccc}
\textbf{I}_{n}& \textbf{X} & \textbf{G}\\
\bm{0}_{p,n} & \textbf{I}_{p} & \bm{0}_{p,r}\\
\bm{0}_{r,n} &  \bm{0}_{r,p} & \textbf{I}_{r}\\
\textbf{I}_{n} &  \bm{0}_{n,p} & \bm{0}_{n,r}
\end{array}
\right),
\end{equation}
\noindent
so that $\textbf{Q}\textbf{Q}^{\prime} = \textbf{I}_{2n+p+r} - \textbf{H}(\textbf{H}^{\prime}\textbf{H})^{-1}\textbf{H}^{\prime}$ and $\textbf{H}^{\prime}\textbf{Q} = \bm{0}_{n+p+r,n}$, where recall that idempotent matrices have eigenvalues equal to zero or one. Let $\textbf{D}(\bm{\theta})^{-1} = blkdiag(\textbf{I}_{n},\textbf{D}_{\beta}(\bm{\theta})^{-1},\textbf{D}_{\eta}(\bm{\theta})^{-1},\frac{1}{\sigma_{\xi}^{2}}\textbf{I}_{n})$, where ``blkdiag'' be the block diagonal operator. The free parameter $\textbf{q}$ is now referred as the ``discrepancy term,'' which is assumed unknown. Several LGPs in the literature set $\textbf{q} = \bm{0}_{n,1}$. However, if one instead assumes an improper prior on $\textbf{q}$ then the posterior distribution of $\bm{\zeta} = (\bm{\xi}^{\prime},\bm{\beta}^{\prime},\bm{\eta}^{\prime})^{\prime}$ and $\textbf{q}$ is GCM, as seen in Theorem~\ref{thm:4} below. {In Appendix C, we derive the posterior distribution up to a proportionality constant when these discrepancy parameters are set equal zero, which can not be simulated from directly.}

\noindent
\begin{theorem}\label{thm:4}
Suppose $Z_{i}\vert Y_{i},b_{i,k}$ are independently distributed according to (\ref{EF2}). For $k = 1$, let the prior for $\sigma_{i}^{2} \in \bm{\theta}$ be an inverse gamma distribution with shape $\alpha_{\sigma}-0.5$ and scale $\kappa_{\sigma}$. Assume the model for $\textbf{y}$ in (\ref{mem2}), the improper prior $f(\textbf{q}) = 1$, $\textbf{D}(\bm{\theta})^{-1} = blkdiag(\textbf{I}_{n},\textbf{D}_{\beta}(\bm{\theta})^{-1},\textbf{D}_{\eta}(\bm{\theta})^{-1},\frac{1}{\sigma_{\xi}^{2}}\textbf{I}_{n})$, and let the hyperparameters $\bm{\theta}$ have a proper prior distribution $\pi(\bm{\theta}) = \pi(\bm{\theta}\cap \{\sigma_{i}^{2}\}^{c})\prod_{i = 1}\pi(\sigma_{i}^{2})$ where ``c'' denotes the set complement. Let $\pi_{*}(\bm{\theta}) =\pi(\bm{\theta}\cap \{\sigma_{i}^{2}\}^{c})\prod_{i = 1}\pi_{*}(\sigma_{i}^{2})$  with $\pi_{*}(\sigma_{i}^{2})$ inverse gamma with shape $\alpha_{\sigma}$ and scale $\kappa_{\sigma}$. Then
	\begin{equation*}
(\bm{\zeta}^{\prime},\textbf{q}^{\prime})^{\prime}\vert \textbf{z}\sim \mathrm{GCM}(\bm{\alpha}_{M},\bm{\kappa}_{M}, \bm{0}_{2n+p+r,1},\textbf{V}_{M},\pi_{*},\textbf{D}; \bm{\psi}_{M}),
	\end{equation*} 
	 where $\textbf{V}_{M}^{-1} = (\textbf{H},\textbf{Q})$ is defined by (\ref{standardH}), $\textbf{D}_{\sigma} = \mathrm{diag}\left(\frac{1}{\sigma_{i}^{2}}: i = 1,\ldots, n\right)$, the $(2n+p+r)$-dimensional unit-log partition function $\bm{\psi}_{M}(\textbf{h}) =\left(\psi_{k}(h_{1}),\ldots,\psi_{k}(h_{n}),\ldots,\psi_{1}(h_{1}^{*}),\ldots, \psi_{1}(h_{n+p+r}^{*})\right)^{\prime}$ for $(2n+p+r)$-dimensional real-valued vector $\textbf{h}= (h_{1},\ldots, h_{n}, h_{1}^{*},\ldots, h_{n+p+r}^{*})^{\prime}$,  and the $(2n+p+r)$-dimensional location and shape/scale parameter vectors are defined as follows: $\bm{\alpha}_{M} = (\textbf{z}^{\prime}\textbf{D}_{\sigma}^{\prime},\bm{0}_{1,n})^{\prime}$ and $\bm{\kappa}_{M} = \frac{1}{2}\bm{1}_{2n,1}$ when the data is normally distributed; $\bm{\alpha}_{M} = (\textbf{z}^{\prime}+\alpha_{\xi}\bm{1}_{1,n},\bm{0}_{1,n})^{\prime}$ and $\bm{\kappa}_{M} = (\bm{1}_{1,n},\frac{1}{2}\bm{1}_{1,n})^{\prime}$ when the data is Poisson distributed; and $\bm{\alpha}_{M} = (\textbf{z}^{\prime}+\alpha_{\xi}\bm{1}_{1,n},\bm{0}_{1,n})^{\prime}$ and $\bm{\kappa}_{M} = (\textbf{m}^{\prime}+2\alpha_{\xi}\bm{1}_{1,n},\frac{1}{2}\bm{1}_{1,n})^{\prime}$ when the data is binomial distributed.
\end{theorem}
\noindent
\textit{Proof:} See Appendix B.\\

\noindent
In Theorem~\ref{thm:4} the presence of $\alpha_{\xi}>0$ make components of $\bm{\alpha}_{M}$ and $\bm{\kappa}_{M}$ strictly positive when elements of the Poisson or binomial data vectors $\textbf{z}$ are zero-valued. Hence, the presence of a cGCM (chosen to be close to a Gaussian) fine-scale term allows one to avoid the boundaries of the parameter space, leading to a well-defined GCM. Theorem~\ref{thm:4} allows one to obtain replicates directly from the posterior distribution $f(\bm{\zeta},\textbf{q}\vert \textbf{z})$ using a familiar projection expression, as seen below in Theorem~\ref{cor:2}.\\

\begin{theorem}\label{cor:2}
{Denote a replicate of $\bm{\zeta}$, $\textbf{q}$, and $\textbf{y}$ using $f(\bm{\zeta},\textbf{q}\vert \textbf{z})$ from Theorem~(\ref{thm:4}) with $\bm{\zeta}_{rep}$, $\textbf{q}_{rep}$, and $\textbf{y}_{rep}$. Then }
\begin{align}\label{zetasim}
\bm{\zeta}_{rep} &=(\textbf{H}^{\prime}\textbf{H})^{-1}\textbf{H}^{\prime}\textbf{w}\\
\label{qsim}
\textbf{q}_{rep} &=\textbf{Q}^{\prime}\textbf{w},\\
\label{ysim}
\textbf{y}_{rep} &=(\textbf{I}_{n},\bm{0}_{n,n+p+r})\textbf{H}\bm{\zeta}_{rep}+(\textbf{I}_{n},\bm{0}_{n,n+p+r})\textbf{Q}\textbf{q}_{rep}= (\textbf{I}_{n},\bm{0}_{n,n+p+r})\textbf{w}
\end{align}
\noindent
\textit{where the $(2n+p+r)$-dimensional random vector $\textbf{w}$ is GCM$(\bm{\alpha}_{M},\bm{\kappa}_{M},\bm{0}_{2n+p+r,1},\textbf{I}_{2n+p+r}, \pi_{*}, \textbf{D}; \bm{\psi}_{M})$, where $\bm{\alpha}_{M}$, $\bm{\kappa}_{M}$, $\pi_{*}$, $\textbf{D}$, and $\bm{\psi}_{M}$ are the same as defined in Theorem~\ref{thm:4}.}
\end{theorem}
\noindent
\textit{Proof:} See Appendix A\\

\noindent
The vector $(2n+p+r)$-dimensional vector $\textbf{w}\equiv\textbf{D}(\bm{\theta})\textbf{w}_{M} = (\textbf{y}_{rep}^{\prime},\textbf{w}_{\beta}^{\prime},\textbf{w}_{\eta}^{\prime},\textbf{w}_{\xi}^{\prime})^{\prime}$, where $\textbf{y}_{rep}$ is easy to generate since it consists of independent DY random variables, $\textbf{w}_{\beta}\sim f(\bm{\beta}\vert \bm{\alpha}_{\beta},\bm{\kappa}_{\beta},\bm{\delta}_{\beta} = \bm{0}_{p,1})$ is $p$-dimensional, $\textbf{w}_{\eta}\sim f(\bm{\eta}\vert \bm{\alpha}_{\beta},\bm{\kappa}_{\beta},\bm{\delta}_{\eta} = \bm{0}_{r,1})$ is $r$-dimensional, and $\textbf{w}_{\xi}$ is $n$-dimensional consisting of independent Gaussian random variables with mean zero and variance $\sigma_{\xi}^{2}$. Note that $\textbf{w}_{\beta}$ and $\textbf{w}_{\eta}$ are simply samples from the respective marginal prior distributions for $\bm{\beta}$ and $\bm{\eta}$ after marginalizing across $\bm{\theta}$ and with location vector zero. Thus, it is straightforward to compute $\textbf{w}$ when it is straightforward to sample from the marginal prior distributions for $\bm{\beta}$ and $\bm{\eta}$. To do this one can, for example, sample from the joint distribution of $\bm{\beta}$ and $\bm{\theta}$, where first one samples $\bm{\theta}$ from $\pi$ then samples from $f(\bm{\beta}\vert \bm{\alpha}_{\beta}, \bm{\kappa}_{\beta},\bm{\delta}_{\beta} = \bm{0}_{p,1}, \textbf{D}_{\beta}(\bm{\theta}))$. The projection $(\textbf{H}^{\prime}\textbf{H})^{-1}\textbf{H}^{\prime} \textbf{w}$ can be computed on the order of $n+p^{3}+r^{3}$ operations with storage on the order of $n(p+r)+p^{2}+r^{2}$, when $\textbf{G}$ is dense. When $\textbf{G}$ is identity with $r = n$, $(\textbf{H}^{\prime}\textbf{H})^{-1}\textbf{H}^{\prime} \textbf{w}$ can be computed on the order of $p^{3}$ operations with storage on the order of $np+p^{2}$. For the details on computing $(\textbf{H}^{\prime}\textbf{H})^{-1}\textbf{H}^{\prime} \textbf{w}$ see Section~\ref{sec:comp}.

The solution in the Gaussian special case is very similar to that in \citet{murphy2007conjugate} and \citet{zhang2023exact}. Namely, a different regression arises in (\ref{zetasim}) from \citet{murphy2007conjugate} and \citet{zhang2023exact} due to our incorporation of a fine-scale variability term. Recall that the presence of fine-scale terms is particularly important for non-Gaussian data, since shape parameters and rate parameters in $\bm{\alpha}_{M}$ and $\bm{\kappa}_{M}$ are non-zero when count-valued observations are zero (i.e., the first stack components of $\bm{\alpha}_{M}$ and $\bm{\kappa}_{M}$) leading to a proper GCM. Thus, one exciting feature of Theorem~\ref{cor:2} is that we obtain Gaussian like simulations of posterior replicates from the posterior distribution for non-Gaussian data. Equation (\ref{zetasim}) can also be seen as a parsimonious special case of the sampler in \citet{bradley2023deep} with considerably fewer parameters. 

 Theorem~\ref{cor:2} provides the motivation for including the discrepancy parameter $\textbf{q}$. Namely, this discrepancy parameter leads to easy-to-compute direct simulations from the posterior distribution. However, the incorporation of $\textbf{q}$ leads to a model that is clearly overparameterized. Thus, a simple solution is to perform inference on $\bm{\zeta}$ using exact replicates from (\ref{zetasim}), which generates values from the marginal distribution $f(\bm{\zeta}\vert  \textbf{z})$. Then use the estimator of $\textbf{q}=\bm{0}_{n,1}$. This is the general strategy used in the CM literature \citep{bradleyLCM} implemented using a type of block Gibbs sampler. Let $\widehat{\textbf{y}}$ represent the profile of $\textbf{y}$ using the plug-in estimator $\textbf{q} = \bm{0}_{n,1}$, so that $\widehat{\textbf{y}}_{rep} =(\textbf{I}_{n},\bm{0}_{n,n+p+r})\textbf{H}\bm{\zeta}_{rep}= \textbf{X}\bm{\beta}_{rep}+\textbf{G}\bm{\eta}_{rep} + \bm{\xi}_{rep}$, where $\bm{\zeta}_{rep} = (\bm{\xi}_{rep}^{\prime},\bm{\beta}_{rep}^{\prime},\bm{\eta}_{rep}^{\prime})^{\prime}$. Moreover, one might similarly use $\widetilde{\textbf{y}}_{rep} = \textbf{X}\bm{\beta}_{rep}+\textbf{G}\bm{\eta}_{rep}$ for inference on $\textbf{y}$, which would implicitly estimate both $\textbf{q}$ and $\bm{\xi}$ to be zero after marginalizing them from the posterior distribution.
 
  The random vector $\textbf{y}_{rep}$ has a very important interpretation. If one assumes $Z_{i}\vert Y_{i}$ is distributed according to the natural exponential family in (\ref{EF2}), and $Y_{i}$ is independently distributed according to the DY distribution in (\ref{univ_LG}) then we have that the implied posterior distribution for $\{Y_{i}\}$ in (\ref{saturated}) is equal in distribution to $\textbf{y}_{rep}$ in Theorem~\ref{cor:2}. Thus, $\textbf{y}_{rep}$ represents a replicate from the posterior distribution from a saturated model. Recall in the goodness-of-fit literature that saturated models define a separate parameter for each datum and is meant to overfit the data, and then, measures of deviance from the saturated model are used to select more parsimonious models \citep[e.g., see][for a recent paper]{bradleyHGT}. This provides additional motivation for using the marginal distribution $f(\bm{\zeta}\vert \textbf{z})$ and $\widehat{\textbf{y}}_{rep}$ (or $\widetilde{\textbf{y}}_{rep}$) to perform inference on $\textbf{y}$, which implies the use of the estimator of $\textbf{q}=\bm{0}_{n,1}$ (and $\bm{\xi}=\bm{0}_{n,1}$). In the recent literature  $\textbf{y}_{rep}-\widehat{\textbf{y}}_{rep}$ $(= -\bm{\delta})$ is referred to as ``discrepancy error,'' and hence we refer to $\bm{\delta}$ as a discrepancy term \citep{bradleyhierarchical,bradley2023deep}.

\subsection{Spatial Process Modeling with Exact Posterior Regression}
\label{sec:process}The mixed effects model specification in Section~4.1 may be deceptively simple; however, we emphasize that several modern statistical models can use EPR including process models (e.g., spatial and spatio-temporal statistical models). See \citet{hodges2013richly} for an thorough  treatment of how spatial and temporal statistical models can be written as a richly parameterized mixed effects model. Although, of course, process models are different from mixed effects models, implementation of additive process models are similar to that of mixed effects models for a given collection of location/times. For example, consider locations $\textbf{s}\in D$, where $D$ is a generic spatial domain (e.g., a lattice or subset of $\mathbb{R}^{d}$). We introduce process into our notation functionally so that, for example, $Z_{i}$ is written as $Z(\textbf{s}_{i})$, where $\textbf{s}_{1},\ldots, \textbf{s}_{n} \in D$. Consider the following multivariate spatial statistical model,
\begin{equation*}
	Y(\textbf{s}) = \textbf{x}(\textbf{s})^{\prime}\bm{\beta} + \textbf{g}(\textbf{s})^{\prime}\bm{\eta} + (\xi(\textbf{s}) - \delta(\textbf{s})); \hspace{2pt}\textbf{s} \in D,
\end{equation*}
where $\textbf{x}(\textbf{s})$ is a $p$-dimensional vector of spatially varying covariates, $\textbf{g}(\textbf{s})$ is a $r$-dimensional vector of spatial basis functions, $\xi(\textbf{s})$ be a random process, and $\delta(\textbf{s})$ be an unknown mean function. Suppose we are interested in estimation and prediction at the observed locations $D_{O}=\{\textbf{s}_{i}: i = 1,\ldots, n\}$ and an additional $m$ locations $D_{P}\in \{\textbf{u}_{1},\ldots, \textbf{u}_{m}\}\subset D$. Let $M = n+m$. Then stacking over locations in $D_{O}\cup D_{P}$ yields,
\begin{equation}\label{process}
	\textbf{y} = \textbf{X}\bm{\beta} + \textbf{G}\bm{\eta} + (\bm{\xi}-\bm{\delta}_{y}),
\end{equation} 
\noindent
where ``$\cup$'' is the set union, $n$-dimensional vector $\textbf{y} = (Y(\textbf{s}): \textbf{s}\in D_{O})^{\prime}$, and the $n\times p$ matrix $\textbf{X}=(\textbf{x}(\textbf{s}): \textbf{s}\in D_{O})^{\prime}$, where we note that $\textbf{X}$ can be computed by pre-multiplying the covariates stacked over $D_{O}\cup D_{P}$ by a $n\times M$ incidence matrix $\textbf{E} = (\textbf{e}(\textbf{s}): s \in D_{O}\cup D_{P})^{\prime}$ with $M$-dimensional vector $\textbf{e}(\textbf{s}) \equiv (I(\textbf{s} = \textbf{s}_{1}),\ldots I(\textbf{s} = \textbf{s}_{n}),I(\textbf{s} = \textbf{u}_{1}),\ldots, I(\textbf{s} = \textbf{u}_{m}))^{\prime}$ and $I(\cdot)$ denoting the indicator function. That is $\textbf{X} = \textbf{E}\textbf{X}_{M}$, where the $M\times p$ matrix $\textbf{X}_{M}=(\textbf{x}(\textbf{s}): \textbf{s}\in D_{O}\cup D_{P})^{\prime}$. In a similar manner let the $n\times M$ matrix $\textbf{G} = \textbf{E}\textbf{G}_{M}$, where the $M\times M$ matrix $\textbf{G}_{M}=(\textbf{g}(\textbf{s}): \textbf{s}\in D_{O}\cup D_{P})^{\prime}$. Here, we let $\textbf{G}_{M}$ be the matrix square root of a parameterized covariance matrix (e.g., $\textbf{G}_{M}$ may be the Cholesky of a $M\times M$ covariance matrix with $(i,j)$-th element defined by the exponential covariogram). We let $\bm{\xi}$, $\bm{\beta}$, $\bm{\eta}$, and $\bm{\delta}_{y}$ in (\ref{process}) have the same specifications as in Section~\ref{sec:epr} with $\textbf{D}_{\eta}\equiv \textbf{I}_{r}$. Comparing our mixed effects model setup in (\ref{mem2}) and the process model specification in (\ref{process}) we see that process modeling can be implemented with EPR. That is, Theorem~\ref{thm:4} (i.e., EPR) can be applied to the stacked expression in Equation (\ref{process}). We illustrate this with spatial basis function expansions, weakly stationary spatial processes, and the conditional autoregressive model \citep{besagYorkMollie} in Section~\ref{sec:empr}. To predict the process at both observed and prediction locations, posterior summaries of $\widetilde{\bm{y}}_{M} = \textbf{X}_{M}\bm{\beta} + \textbf{G}_{M}\bm{\eta}$ will be used

In practice, EPR may not always be scale-able for process modeling with large $n$ and $M$, since it is not always straightforward to simulate directly from the prior distribution, nor is it always straightforward to compute $\textbf{G}$. In this article, we consider one example with a reduced rank assumption \citep{johan,banerjee,hughes} by defining $\textbf{G}$ to consist of $r<M$ spatially referenced basis functions (e.g., see Section~\ref{sec:sbf}). Although we consider $r<M$ to achieve scalability, there are options to consider when implementing EPR with $r\ge M$. In particular, one might consider the ``data subset model'' from \citep{bradley2021approach} to achieve scale-able inference, or sparse matrix Cholesky decompositions \citep[e.g., see][]{datta2016hierarchical}. 

\subsection{Computational Considerations}\label{sec:comp}

For large $n$ the EPR formulation may not look practically feasible. However, standard block matrix inversion techniques can be used to reduce the order of operations to inverses of $r\times r$ matrices, $p\times p$ matrices, and $n\times n$ diagonal matrices \citep{lu2002inverses}. 

\noindent
\begin{theorem}\label{thm:5}
	\textit{The following expression holds,}
	\begin{equation}\label{firstblock}
		(\textbf{H}^{\prime}\textbf{H})^{-1} = \left(\begin{array}{cc}
			\textbf{A}^{-1} + \textbf{A}^{-1}\textbf{B}(\textbf{D} - \textbf{B}^{\prime}\textbf{A}^{-1}\textbf{B})^{-1}\textbf{B}^{\prime}\textbf{A}^{-1} & -\textbf{A}^{-1}\textbf{B}(\textbf{D} - \textbf{B}^{\prime}\textbf{A}^{-1}\textbf{B})^{-1}\\
			-(\textbf{D} - \textbf{B}^{\prime}\textbf{A}^{-1}\textbf{B})^{-1}\textbf{B}^{\prime}\textbf{A}^{-1} & (\textbf{D} - \textbf{B}^{\prime}\textbf{A}^{-1}\textbf{B})^{-1}
		\end{array}
		\right),
	\end{equation}
	\noindent
	\textit{where $\textbf{A} =2\textbf{I}_{n}$, the $n\times (p+r)$ matrix $\textbf{B} = (\textbf{X},\hspace{2pt}\textbf{G})$, the $(p+r)\times (p+r)$ matrix }
	\begin{equation}\label{secondblock}
		\textbf{D}   = \left(\begin{array}{cc}
			\textbf{X}^{\prime}\textbf{X} + \textbf{I}_{p} & \textbf{X}^{\prime}\textbf{G}\\
			\textbf{G}^{\prime}\textbf{X} & \textbf{G}^{\prime}\textbf{G} + \textbf{I}_{r}
		\end{array}
		\right),
	\end{equation}
	\noindent
	\textit{the $(p+r)\times (p+r)$ matrix } 
	\begin{equation}
		\nonumber
		(\textbf{D} - \textbf{B}^{\prime}\textbf{A}^{-1}\textbf{B})^{-1} = \left(\begin{array}{cc}
			\textbf{A}^{*-1} + \textbf{A}^{*-1}\textbf{B}^{*}(\textbf{D}^{*} - \textbf{C}^{*}\textbf{A}^{*-1}\textbf{B}^{*})^{-1}\textbf{C}^{*}\textbf{A}^{*-1} & -\textbf{A}^{*-1}\textbf{B}^{*}(\textbf{D}^{*} - \textbf{C}^{*}\textbf{A}^{*-1}\textbf{B}^{*})^{-1}\\
			-(\textbf{D}^{*} - \textbf{C}^{*}\textbf{A}^{*-1}\textbf{B}^{*})^{-1}\textbf{C}^{*}\textbf{A}^{*-1} & (\textbf{D}^{*} - \textbf{C}^{*}\textbf{A}^{*-1}\textbf{B}^{*})^{-1}
		\end{array}
		\right),
	\end{equation}
	\noindent
	\textit{the $p\times p$ matrix $\textbf{A}^{*} = \frac{1}{2}\textbf{X}^{\prime}\textbf{X} + \textbf{I}_{p}$, the $p\times r$ matrix $\textbf{B}^{*} = \frac{1}{2}\textbf{X}^{\prime}\textbf{G}$, the $r\times p$ matrix $\textbf{C}^{*} = \frac{1}{2}\textbf{G}^{\prime}\textbf{X}$, and the $r\times r$ matrix $\textbf{D}^{*} = \frac{1}{2}\textbf{G}^{\prime}\textbf{G}+ \textbf{I}_{r}$.}
\end{theorem}
\noindent
\textit{Proof:} See Appendix B.\\

\noindent
Theorem~\ref{thm:5} allows us to reduce the inverse of the $(n+p+r)\times (n+p+r)$ matrix $\textbf{H}^{\prime}\textbf{H}$ to the inverse of the $p\times p$ matrix $\textbf{A}^{*}$, and the $r\times r$ matrix $(\textbf{D}^{*} - \textbf{C}^{*}\textbf{A}^{*-1}\textbf{B}^{*})^{-1}$. When $p$ and $r$ are both ``small,'' these inverses are computationally efficient.

Simulation from the posterior using EPR does not necessarily require first computing a matrix of the form $(\textbf{H}^{\prime}\textbf{H})^{-1}$, storing this matrix, and then computing a $(n+p+r)$-dimensional vector of the form $(\textbf{H}^{\prime}\textbf{H})^{-1}\textbf{H}^{\prime}\textbf{w}$. In fact this order of operations may require impractical storage, since the $(n+p+r)\times (n+p+r)$ matrix $(\textbf{H}^{\prime}\textbf{H})^{-1}$ may be high-dimensional. To avoid these issues one can instead compute/store the $(n+p+r)$-dimensional vector of the form $(\textbf{H}^{\prime}\textbf{H})^{-1}\textbf{H}^{\prime}\textbf{w}$ that avoids storage of high-dimensional matrices.
\begin{theorem}\label{thm:7}
	\textit{Let $\textbf{w} = (\textbf{w}_{e}^{\prime},\textbf{w}_{\beta}^{\prime},\textbf{w}_{\eta}^{\prime},\textbf{w}_{q}^{\prime})^{\prime}$, $\textbf{w}_{e}\in \mathbb{R}^{n}$, $\textbf{w}_{\beta}\in \mathbb{R}^{p}$, $\textbf{w}_{\eta}\in \mathbb{R}^{r}$, and $\textbf{w}_{\xi}\in \mathbb{R}^{n}$. Then the following expression holds,}
	\begin{equation}\label{firstvecrep}
		(\textbf{H}^{\prime}\textbf{H})^{-1}\textbf{H}^{\prime}\textbf{w} = \left(\begin{array}{c}
			(\textbf{F} - \textbf{K}\textbf{L}^{-1}\textbf{K}^{\prime})^{-1}(\textbf{R}-\textbf{K}\textbf{L}^{-1}\textbf{P}) \\
			-\textbf{L}^{-1}\textbf{K}^{\prime}(\textbf{F} - \textbf{K}\textbf{L}^{-1}\textbf{K}^{\prime})^{-1}(\textbf{R}-\textbf{K}\textbf{L}^{-1}\textbf{P}) + \textbf{L}^{-1}\textbf{P}
		\end{array}
		\right),
	\end{equation}
	\noindent
	\textit{where the $(n+p)$-dimensional vector $\textbf{R} = (\textbf{w}_{e}^{\prime}+ \textbf{w}_{q}^{\prime},\textbf{w}_{e}^{\prime}\textbf{X}+ \textbf{w}_{\beta}^{\prime})^{\prime}$, the $r$-dimensional vector $\textbf{P} = \textbf{G}^{\prime}\textbf{w}_{e} + \textbf{w}_{\eta}$, the $r\times (n+p)$ matrix $\textbf{K}^{\prime} = (\textbf{G}^{\prime}, \textbf{G}^{\prime}\textbf{X})$, the $r\times r$ matrix $\textbf{L} = \textbf{G}^{\prime}\textbf{G}+\textbf{I}_{r}$, the $(n+p)\times (n+p)$ matrix $\textbf{F}   = \left(\begin{array}{cc}2\textbf{I}_{n}
			& \textbf{X}\\
			\textbf{X}^{\prime} & \textbf{X}^{\prime}\textbf{X} + \textbf{I}_{p}
		\end{array}
		\right)$, and the $(n+p)\times (n+p)$ matrix}
	\begin{equation}
		\nonumber
		\textbf{F}- \textbf{K}\textbf{L}^{-1}\textbf{K}^{\prime}   = \left(\begin{array}{cc}\textbf{F}_{1}
			& \textbf{B}_{12}\\
			\textbf{B}_{12}^{\prime} & \textbf{F}_{2}
		\end{array}
		\right).
	\end{equation}
	\noindent
	\textit{The $(n+p)\times (n+p)$ matrix,} 
	\begin{equation}
		\nonumber
		(\textbf{F} - \textbf{K}\textbf{L}^{-1}\textbf{K}^{\prime})^{-1}= \left(\begin{array}{cc}
			\textbf{F}_{11} & \textbf{F}_{12}\\
			\textbf{F}_{21} & \textbf{F}_{22}
		\end{array}
		\right),
	\end{equation}
	\noindent
	\textit{where the $n\times n$ matrix $\textbf{F}_{1} = 2\textbf{I}_{n} - \textbf{G}\textbf{L}^{-1}\textbf{G}^{\prime}$, the $n\times p$ matrix $\textbf{B}_{12} = \textbf{X}  - \textbf{G}\textbf{L}^{-1}\textbf{G}^{\prime}\textbf{X}$, the $p\times p$ matrix $\textbf{F}_{2} = \textbf{X}^{\prime}\textbf{X} + \textbf{I}_{p}- \textbf{X}^{\prime}\textbf{G}\textbf{L}^{-1}\textbf{G}^{\prime}\textbf{X}$, the $n\times n$ matrix $\textbf{F}_{11} = \textbf{F}_{1}^{-1} +\textbf{F}_{1}^{-1}\textbf{B}_{12}(\textbf{F}_{2}-\textbf{B}_{12}^{\prime}\textbf{F}_{1}^{-1}\textbf{B}_{12})^{-1}\textbf{B}_{12}^{\prime}\textbf{F}_{1}^{-1}$, the $n\times p$ matrix $\textbf{F}_{12} =- \textbf{F}_{1}^{-1}\textbf{B}_{12}(\textbf{F}_{2}-\textbf{B}_{12}^{\prime}\textbf{F}_{1}^{-1}\textbf{B}_{12})^{-1}$, the $p\times n$ matrix\\ $\textbf{F}_{21} = -(\textbf{F}_{2}-\textbf{B}_{12}^{\prime}\textbf{F}_{1}^{-1}\textbf{B}_{12})^{-1}\textbf{B}_{12}^{\prime}\textbf{F}_{1}^{-1}$, the $p\times p$ matrix $\textbf{F}_{22} = (\textbf{F}_{2}-\textbf{B}_{12}^{\prime}\textbf{F}_{1}^{-1}\textbf{B}_{12})^{-1}$, and the $n\times n$ matrix $\textbf{F}_{1}^{-1} = \frac{1}{2}\textbf{I}_{n} + \frac{1}{4}\textbf{G}(\textbf{L} - \frac{1}{2}\textbf{G}^{\prime}\textbf{G})^{-1}\textbf{G}^{\prime}$.}
\end{theorem}
\noindent
\textit{Proof:} See the Appendix.\\

\noindent
Careful examination of the order of operations show that Theorem~(\ref{thm:7}) allows one to compute the vector $(\textbf{H}^{\prime}\textbf{H})^{-1}\textbf{H}^{\prime}\textbf{w}$ by storing/computing the $n\times p$ matrix $\textbf{X}$, the $n\times r$ matrix $\textbf{G}$ (when $r = n$ we set $\textbf{G} = \textbf{I}_{n}$), the $r\times r$ matrix $\textbf{L}^{-1}$, the $r\times r$ matrix $(\textbf{L} - \frac{1}{2}\textbf{G}^{\prime}\textbf{G})^{-1}$, the $p\times p$ matrix $\textbf{F}_{2}$, and the $p\times p$ matrix $\textbf{F}_{22}$. These computations are straightforward when $r$ and $p$ are ``small'' or when $p$ is small and $\textbf{G}$ is diagonal.

\subsection{Implementation of Exact Posterior Regression}\label{sec:stepbystep}

The following gives step-by-step instructions on obtaining efficient independent replicates directly from the posterior distribution of fixed effects and random effects using Theorem~\ref{thm:4}, which we refer to as EPR. We consider Gaussian data with unknown non-constant variance, Poisson data, binomial data, and Gaussian priors on $\bm{\beta}$ and $\bm{\eta}$. Minor adjustments to these steps are needed for other settings (e.g., jointly modeling just two data types, etc.).

\begin{enumerate}
	\item Store/compute the $n\times p$ matrix $\textbf{X}$, the $n\times r$ matrix $\textbf{G}$, the $r\times r$ matrix $\textbf{L}^{-1}$, the $r\times r$ matrix $(\textbf{L} - \frac{1}{2}\textbf{G}^{\prime}\textbf{G})^{-1}$, the $p\times p$ matrix $\textbf{F}_{2}$, and the $p\times p$ matrix $\textbf{F}_{22}$. 
	\item Simulate $\textbf{w}$ according to Theorem~\ref{cor:2}.
	\item Use the matrices computed in Step 1 and $\textbf{w}$ in Step 2 to compute $\bm{\zeta}_{rep}$, $\textbf{y}_{rep}$, $\widehat{\textbf{y}}_{rep}$, and $\widetilde{\textbf{y}}_{rep}$ according to Theorem~\ref{thm:7}.
	\item Repeat Steps 2$\--$3 $B$ times when $\textbf{G}$ does not consist of unknown parameters. Repeat Steps 1 $\--$ 3 $B$ times when $\textbf{G}$ is parameterized.
\end{enumerate}
\noindent
When using known basis function expansions to define $\textbf{G}$, repeated matrix operations that one might see in a Gibbs sampler are avoided, since matrix inversions are only required \textit{a single time} in Step 1. Additionally, $B$ does not have to be as large as what one requires for an MCMC, since one does not require a burn-in period, thinning, or have concerns of mixing and positive autocorrelations in the MCMC. 

\begin{figure}[t!]
	\begin{center}
		\begin{tabular}{ccc}
			\includegraphics[width=5cm,height=5cm]{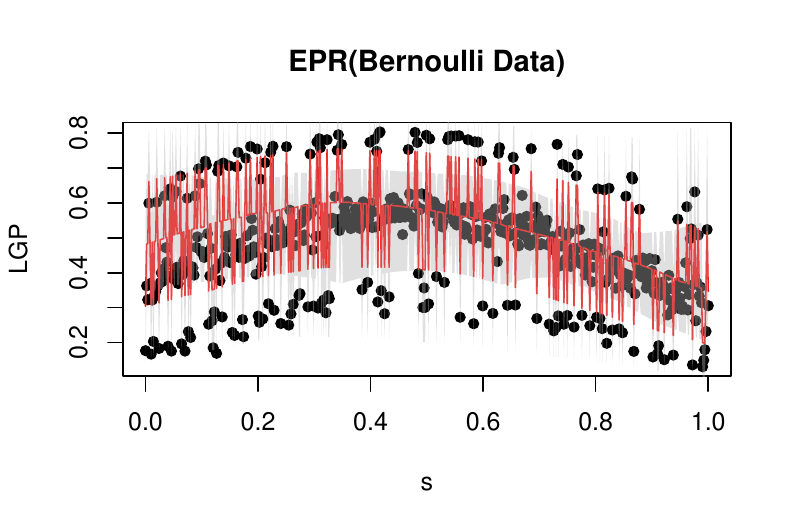}&\includegraphics[width=5cm,height=5cm]{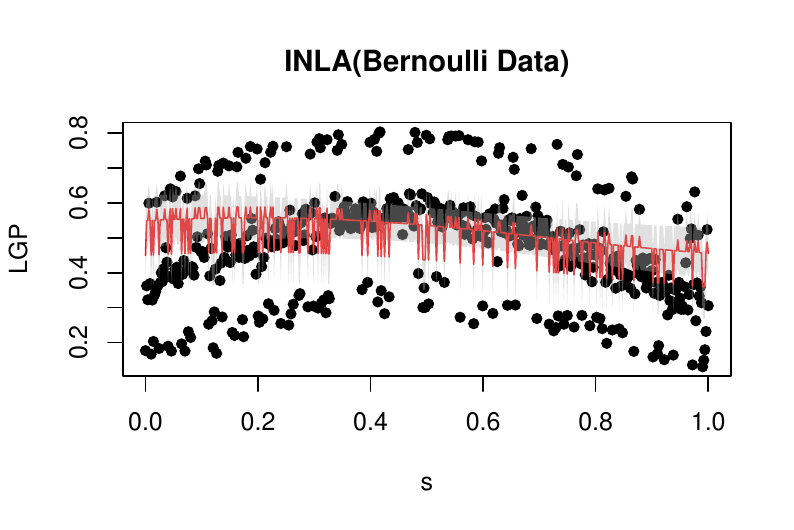}&\includegraphics[width=5cm,height=5cm]{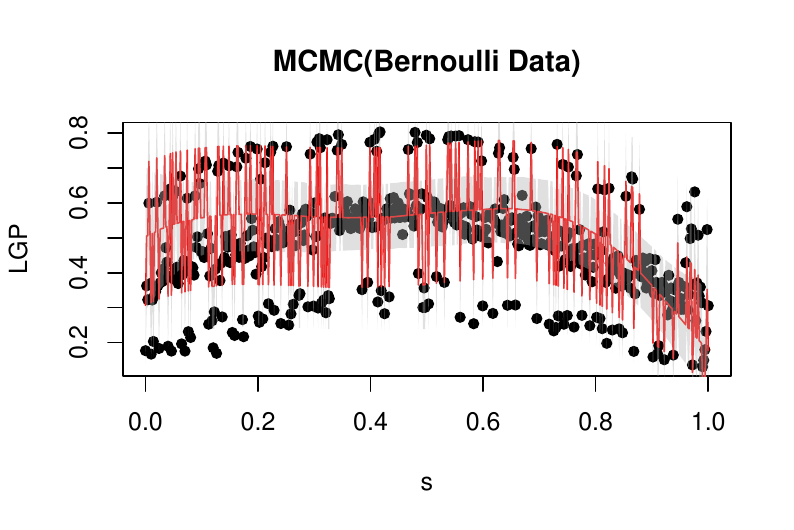}\\
			\includegraphics[width=5cm,height=5cm]{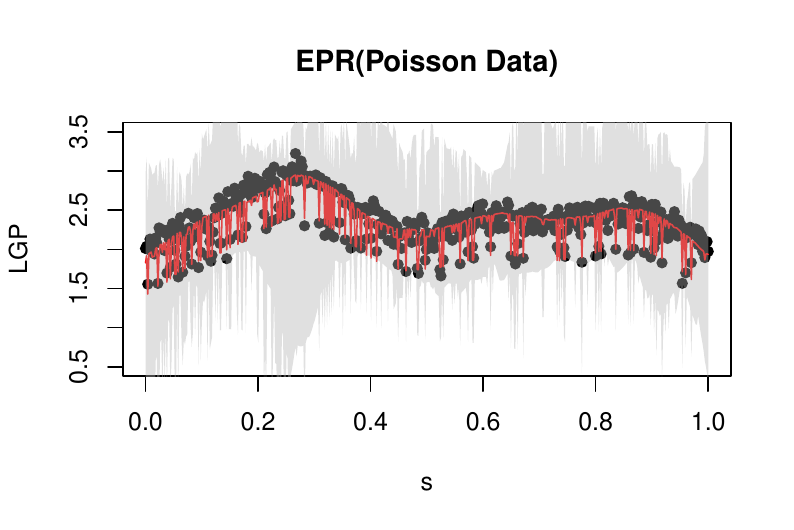}&\includegraphics[width=5cm,height=5cm]{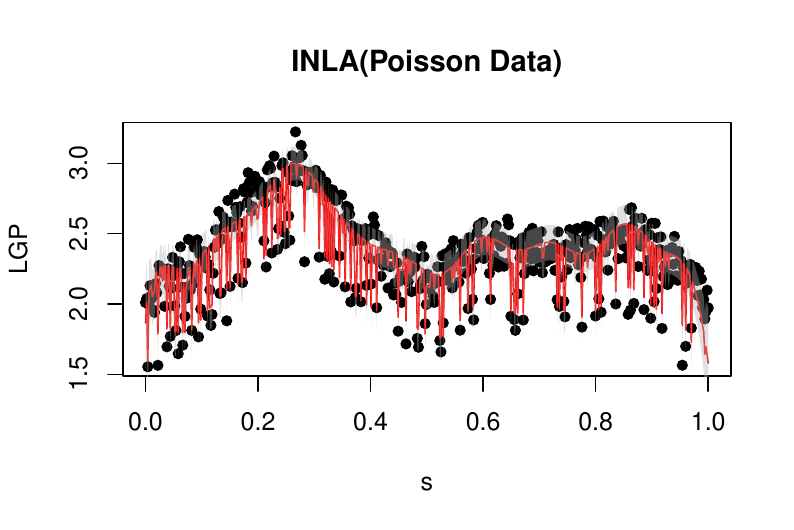}&\includegraphics[width=5cm,height=5cm]{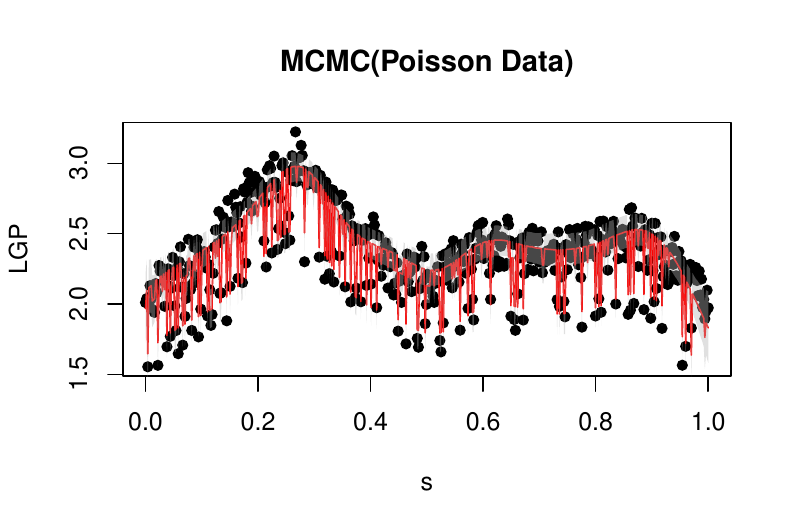}\\
			\includegraphics[width=5cm,height=5cm]{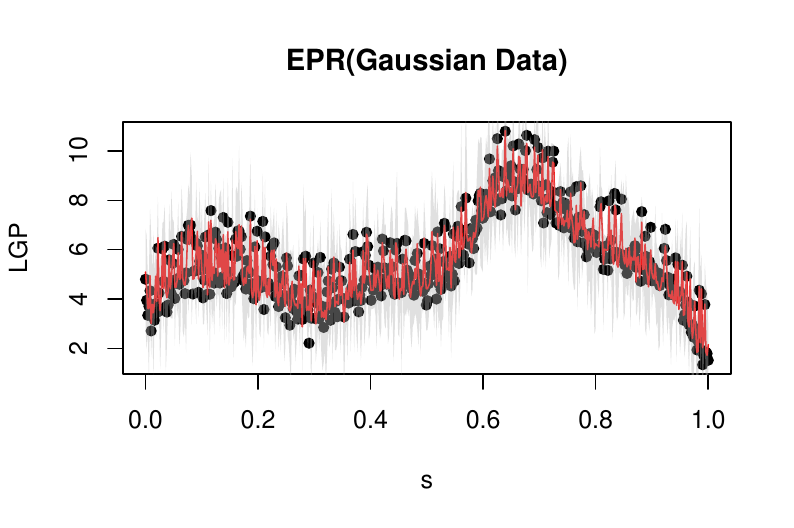}&\includegraphics[width=5cm,height=5cm]{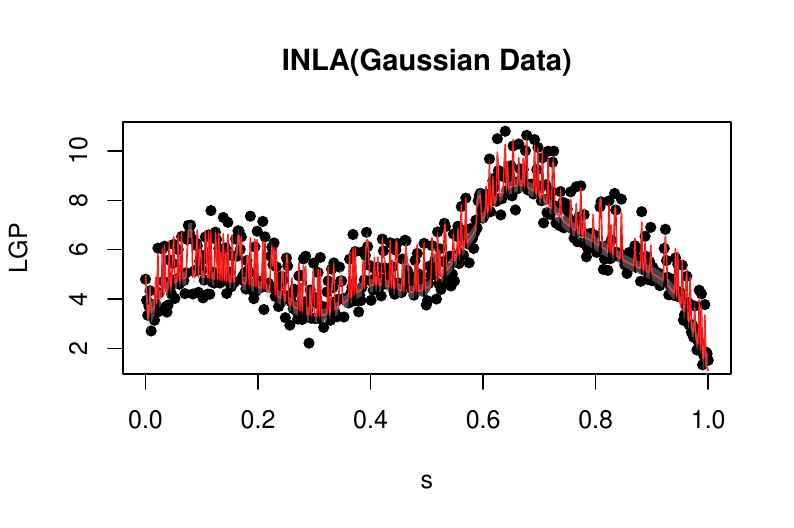}&\includegraphics[width=5cm,height=5cm]{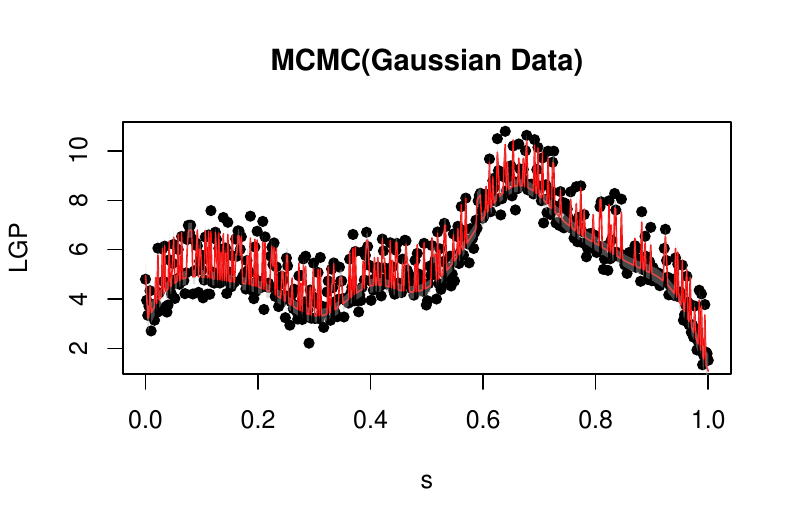}
		\end{tabular}
		
	\end{center}
	\caption{Illustration of EPR, INLA, and MCMC predictions for a spatial basis function expansion. The first row presents results for binary spatial data, the second row presents results for Poisson spatial data, and the third row presents results for Gaussian spatial data. The black points represent the true value of the latent process, the red line represents the posterior mean, and the gray shaded region represents pointwise 95$\%$ credible intervals, respectively.}\label{fig:predexamp2}
\end{figure}

\section{Illustrations}\label{sec:empr}

We provide several illustrations in several standard spatial statistical model settings. In particular, we compare the use of spatial basis function expansions, weakly stationary spatial processes, and conditional autoregressive models, all of which are covered in standard text books on spatial and spatio-temporal modeling \citep[e.g., see][among others]{cressie,Rue2005gaussian,cressie-wikle-book,banerjee-etal-2004}.
%

\subsection{Spatial Basis Function Expansions}\label{sec:sbf} Spatial basis function expansions have become a standard in spatial statistics, with common classes of basis functions including Fourier basis functions, wavelet basis functions \citep{wave-huang}, radial basis functions \citep{johan}, and splines \citep{wahba}, among others. In this section, we compare EPR, INLA, and MCMC implemented using the P\'{o}lya-Gamma technique from \citep{polson,d2022efficient} using Gaussian radial basis function. EPR assumes an inverse gamma prior on all variance parameters with shape 1 and rate parameter given a gamma hyperprior with shape and rate set to 1. The range parameter is given a uniform zero to 0.5 prior. The default prior specifications are used for both INLA and spBayes. The P\'{o}lya-Gamma technique is a particularly efficient approach to fit latent Gaussian process models using MCMC, and is one of the more computationally competitive techniques in MCMC. In the Poisson MCMC implementation we make use of a new extremely efficient algorithm by \citet{d2022efficient}. MCMC was implemented with 3,000 replicates with a burn-in of 1,000. We assume
\begin{align}
	\nonumber
	Z_{1}(\textbf{s})\vert Y(\textbf{s})&\sim \mathrm{Normal}\left(-1-\hspace{2pt}x_{1}(\textbf{s}) -\hspace{2pt}x_{2}(\textbf{s})+ \sum_{j = 1}^{30}g_{j}(\textbf{s})\eta_{j},0.3\right)\\
	\nonumber
	Z_{2}(\textbf{s})\vert Y(\textbf{s})&\sim \mathrm{Poisson}\left\lbrace \mathrm{exp}\left(-1+0.5\hspace{2pt}x_{1}(\textbf{s}) +0.4\hspace{2pt}x_{2}(\textbf{s})+ \sum_{j = 1}^{30}g_{j}(\textbf{s})\eta_{j}\right)\right\rbrace\\
	\label{eq:study2}
	Z_{3}(\textbf{s})\vert Y(\textbf{s})&\sim \mathrm{Bernoulli}\left\lbrace \frac{\mathrm{exp}\left(-2-\hspace{2pt}x_{1}(\textbf{s}) -2\hspace{2pt}x_{2}(\textbf{s})+  \sum_{j = 1}^{30}g_{j}(\textbf{s})\eta_{j}\right)}{1+\mathrm{exp}\left(-2-\hspace{2pt}x_{1}(\textbf{s}) -2\hspace{2pt}x_{2}(\textbf{s}) + \sum_{j = 1}^{30}g_{j}(\textbf{s})\eta_{j}\right)}\right\rbrace,
\end{align}
\noindent
where $\{\eta_{j}\}$ are independently distributed according to a normal distribution with mean zero and variance 0.04, ${s}\in \{0,0.002,\ldots, 1\}$, we observe $n = 400$ randomly selected locations, $x_{1}({s})$ is an independent Bernoulli random variable with probability $\mathrm{exp}({s})/(1+\mathrm{exp}({s}))$, $x_{2}(\textbf{s})$ is an independent Bernoulli random variable with probability $\mathrm{exp}(-0.01 {s})/(1+\mathrm{exp}(-0.01 {s}))$,  $g_{j}(\textbf{{s})}) = \mathrm{exp}(-||{s}-{u}_{j}||^{2})$, $\{{u}_{j}\}$ are equally spaced across the spatial domain, and $||\cdot||$ is the Euclidean distance. The default prior specifications are used for INLA. In Figure~\ref{fig:predexamp2}, we see that each method is fairly comparable in terms of predictive performance, expect in the case of Bernoulli data, where EPR is preferable for this particular dataset. Moreover, EPR tends to give larger measures of uncertainty than INLA and MCMC, both of which produce credible intervals that do not contain the true values of the latent process. The fact that the predictions are similar is notable since, INLA and MCMC are both approximate methods (MCMC is exact in the limit), whereas EPR is an exact method. Moreover, EPR (and INLA) does not require the additional overhead of MCMC diagnostics. 

To assess the performance over multiple replicates, we use the central processing unit (CPU) time (seconds), the mean squared error (MSE) between the estimated regression coefficients and $\{\eta_{j}\}$ and true values, the mean squared prediction error (MSPE) between the latent process and predicted latent process (using $\widetilde{\textbf{y}}$), and the continuous rank probability score (CRPS) \citep{gneiting2014probabilistic} averaged over missing locations and scaled so that small values are preferable. The CRPS is useful since it is metric that evaluates the entire predictive distribution so that uncertainty in the predictions considered. In Table~\ref{tab:comperform2}, we provide the average MSPE, MSE, CRPS, and CPU plus or minus two standard deviations over 50 independent replicates by method and data type.  
\begin{table}[t!]
	\Tiny{
		\begin{center}
			\begin{tabular}{@{} l *5c @{}}
				\toprule
				\multicolumn{1}{c}{Method} & Type  & MSPE & MSE & CRPS & CPU \\ 
				\midrule
				EPR & Logistic & \begin{tabular}{@{}c@{}}0.0037 \\ $(0.0033,0.0042)$\end{tabular} & \begin{tabular}{@{}c@{}}0.245 \\ $(0.168,0.322)$\end{tabular}& \begin{tabular}{@{}c@{}}0.549 \\ $(0.536,0.561)$\end{tabular}& \begin{tabular}{@{}c@{}}0.756  \\ $(0.700,0.813)$\end{tabular}  \\ \hline
				INLA & Logistic & \begin{tabular}{@{}c@{}}0.0084 \\ $(0.0071,0.0096)$\end{tabular} & \begin{tabular}{@{}c@{}}1.259 \\ $(0.712,1.805)$\end{tabular}& \begin{tabular}{@{}c@{}}0.566 \\ $(0.539,0.593)$\end{tabular}& \begin{tabular}{@{}c@{}}4.023 \\ $(3.640,4.407)$\end{tabular} 
				\\ \hline
				MCMC & Logistic & \begin{tabular}{@{}c@{}}0.0047 \\ $(0.0042,0.0052)$\end{tabular}& \begin{tabular}{@{}c@{}}8.345 \\ $(6.189,10.502)$\end{tabular}& \begin{tabular}{@{}c@{}}0.713 \\ $(0.708,0.718)$\end{tabular}& \begin{tabular}{@{}c@{}}58.128 \\ $(54.584,61.673)$\end{tabular} 
				\\ \hline
				EPR & Poisson & \begin{tabular}{@{}c@{}}0.0146 \\ $(0.0135,0.0158)$\end{tabular} & \begin{tabular}{@{}c@{}}0.673 \\ $(0.0491,0.0855)$\end{tabular}& \begin{tabular}{@{}c@{}}0.255 \\ $(0.248,0.263)$\end{tabular}& \begin{tabular}{@{}c@{}}0.47  \\ $(0.437,0.503)$\end{tabular}  \\ \hline
				INLA & Poisson & \begin{tabular}{@{}c@{}}0.0136 \\ $(0.0131,0.0141)$\end{tabular} & \begin{tabular}{@{}c@{}}23.617 \\ $(18.986,28.248)$\end{tabular}& \begin{tabular}{@{}c@{}}0.301 \\ $(0.299,0.303)$\end{tabular}& \begin{tabular}{@{}c@{}}2.068 \\ $(2.042,2.094)$\end{tabular} 
				\\ \hline
				MCMC & Poisson & \begin{tabular}{@{}c@{}}0.0126 \\ $(0.0119,0.0133)$\end{tabular}& \begin{tabular}{@{}c@{}}1.3$\times 10^{10}$ \\ $(2.1\times 10^{9},2.4\times 10^{10})$\end{tabular}& \begin{tabular}{@{}c@{}}8.722\\ $(8.658,8.787)$\end{tabular}& \begin{tabular}{@{}c@{}}23.221 \\ $(23.0178,23.425)$\end{tabular} 
				\\ \hline
				EPR & Normal & \begin{tabular}{@{}c@{}}0.173 \\ $(0.164,0.182)$\end{tabular} & \begin{tabular}{@{}c@{}}1.794\\ $(1.748,1.840)$\end{tabular}& \begin{tabular}{@{}c@{}}1.625 \\ $(1.609,1.641)$\end{tabular}& \begin{tabular}{@{}c@{}}0.48  \\ $(0.44,0.51)$\end{tabular}  \\ \hline
				INLA & Normal & \begin{tabular}{@{}c@{}}0.156 \\ $(0.155,0.157)$\end{tabular} & \begin{tabular}{@{}c@{}}28.325 \\ $(23.766,32.883)$\end{tabular}& \begin{tabular}{@{}c@{}}1.861  \\ $(1.856,1.866)$\end{tabular}& \begin{tabular}{@{}c@{}}2.35 \\ $(2.31,2.39)$\end{tabular} 
				\\ \hline
				MCMC & Normal & \begin{tabular}{@{}c@{}}0.156 \\ $(0.155,0.157)$\end{tabular}& \begin{tabular}{@{}c@{}}26.498 \\ $(22.341,30.656)$\end{tabular}& \begin{tabular}{@{}c@{}}1.860  \\ $(1.855,1.865)$\end{tabular}& \begin{tabular}{@{}c@{}}4.66 \\ $(4.57,4.75)$\end{tabular} \\ \bottomrule
			\end{tabular}
		\end{center}
	}
	\caption{Fifty independent replicates data vectors are drawn according to (\ref{eq:study2}), and several methods are applied to each replicated data vector. The method column indicates EPR, INLA, and HMC. The type column indicates logistic regression, Poisson regression, and normal regression. The values represent averages over 50 independent simulated data sets and the parenthetical represent the confidence interval (CI) (i.e., average plus or minus two standard deviations). The MSE, MSPE, CRPS, and CPU (in seconds) are indicated in the column header. The MSPE for Poisson regression is computed on the log-scale  so that the values are easier to present, where logistic spatial regression's MSPE was computed on the expit scale. }\label{tab:comperform2}
\end{table}

In Table~\ref{tab:comperform2}, we see that the MSPE and CRPS are comparable in magnitude for EPR, INLA, and MCMC. In general, for EPR is preferable in terms of MSPE in the logistic regression setting, and INLA and MCMC produces smaller MSPE for Poisson and Normal regression. EPR and INLA have smaller CRPS (with CI that overlap) than MCMC in the logistic regression setting, and EPR is preferable in terms of CRPS than INLA and MCMC for Poisson and Normal data. EPR is consistently and considerably preferable in terms of MSE and CPU time in all settings. The performance in CPU time is especially notable, since INLA and MCMC are both approximate methods (MCMC is exact in the limit), whereas EPR is an exact MCMC free method. That is, EPR produces comparable predictions and superior regression estimates in a faster time than that of the state-of-the-art approximate Bayes and MCMC based methods in this study.

\begin{table}[t!]
	\Tiny{
	\begin{center}
		\begin{tabular}{@{} l *5c @{}}
			\toprule
			\multicolumn{1}{c}{Method} & Type  & MSPE & MSE & CRPS & CPU \\ 
			\midrule
			EPR & Logistic & \begin{tabular}{@{}c@{}}0.0286 \\ $(0.0273,0.030)$\end{tabular} & \begin{tabular}{@{}c@{}}0.545 \\ $(0.117,0.973)$\end{tabular}& \begin{tabular}{@{}c@{}}1.036 \\ $(0.975,1.097)$\end{tabular}& \begin{tabular}{@{}c@{}}14.1  \\ $(13.8,14.4)$\end{tabular}  \\ \hline
			INLA & Logistic & \begin{tabular}{@{}c@{}}0.030 \\ $(0.0286,0.0317)$\end{tabular} & \begin{tabular}{@{}c@{}}0.343 \\ $(0.247,0.440)$\end{tabular}& \begin{tabular}{@{}c@{}}1.020 \\ $(0.972,1.068)$\end{tabular}& \begin{tabular}{@{}c@{}}8.7 \\ $(5.7,11.7)$\end{tabular} 
			\\ \hline
			MCMC & Logistic & \begin{tabular}{@{}c@{}}0.0387 \\ $(0.0361,0.0412)$\end{tabular}& \begin{tabular}{@{}c@{}}0.363 \\ $(0.250,0.476)$\end{tabular}& \begin{tabular}{@{}c@{}}0.979 \\ $(0.939,1.020)$\end{tabular}& \begin{tabular}{@{}c@{}}344 \\ $(339,349)$\end{tabular} 
			\\ \hline
						EPR & Poisson & \begin{tabular}{@{}c@{}}0.206 \\ $(0.193,0.220)$\end{tabular} & \begin{tabular}{@{}c@{}}2.88 \\ $(2.69,3.07)$\end{tabular}& \begin{tabular}{@{}c@{}}1.22 \\ $(1.16,1.28)$\end{tabular}& \begin{tabular}{@{}c@{}}14.8  \\ $(13.8,15.8)$\end{tabular}  \\ \hline
			INLA & Poisson & \begin{tabular}{@{}c@{}}0.173 \\ $(0.163,0.184)$\end{tabular} & \begin{tabular}{@{}c@{}}1.27 \\ $(0.762,1.77)$\end{tabular}& \begin{tabular}{@{}c@{}}1.34  \\ $(1.25,1.44)$\end{tabular}& \begin{tabular}{@{}c@{}}4.7 \\ $(4.5,4.9)$\end{tabular} 
			\\ \hline
			MCMC & Poisson & \begin{tabular}{@{}c@{}}0.156 \\ $(0.147,0.165)$\end{tabular}& \begin{tabular}{@{}c@{}}2.23 \\ $(1.38,3.08)$\end{tabular}& \begin{tabular}{@{}c@{}}1.42 \\ $(1.32,1.51)$\end{tabular}& \begin{tabular}{@{}c@{}}351.7 \\ $(346.9,354.5)$\end{tabular} 
				\\ \hline
			EPR & Normal & \begin{tabular}{@{}c@{}}0.266 \\ $(0.254,0.279)$\end{tabular} & \begin{tabular}{@{}c@{}}0.358\\ $(0.254,0.462)$\end{tabular}& \begin{tabular}{@{}c@{}}1.078 \\ $(0.963,1.19)$\end{tabular}& \begin{tabular}{@{}c@{}}14.7  \\ $(13.4,15.9)$\end{tabular}  \\ \hline
			INLA & Normal & \begin{tabular}{@{}c@{}}0.244 \\ $(0.229,0.259)$\end{tabular} & \begin{tabular}{@{}c@{}}1.158 \\ $(0.882,1.435)$\end{tabular}& \begin{tabular}{@{}c@{}}1.056  \\ $(1.002,1.111)$\end{tabular}& \begin{tabular}{@{}c@{}}7.9 \\ $(7.3,8.5)$\end{tabular} 
			\\ \hline
			MCMC & Normal & \begin{tabular}{@{}c@{}}0.247 \\ $(0.233,0.262)$\end{tabular}& \begin{tabular}{@{}c@{}}1365.766 \\ $(1238.966,1492.566)$\end{tabular}& \begin{tabular}{@{}c@{}}1.079 \\ $(1.025,1.133)$\end{tabular}& \begin{tabular}{@{}c@{}}112.67 \\ $(106.77,118.57)$\end{tabular} \\ \bottomrule
		\end{tabular}
	\end{center}
}
	\caption{Fifty independent replicates data vectors are drawn according to (\ref{eq:study1}), and several methods are applied to each replicated data vector. The method column indicates EPR, INLA, and HMC. The type column indicates logistic regression, Poisson regression, and normal regression. The values represent averages over 50 independent simulated data sets and the parenthetical represent the confidence interval (CI) (i.e., average plus or minus two standard deviations). The MSE, MSPE, CRPS, and CPU (in seconds) are indicated in the column header. The MSPE for Poisson regression is computed on the log-scale  so that the values are easier to present, where logistic spatial regression's MSPE was computed on the expit scale. }\label{tab:comperform}
\end{table}

	\begin{figure}[t!]
	\begin{center}
		\begin{tabular}{cccc}
			\includegraphics[width=3.5cm,height=3.5cm]{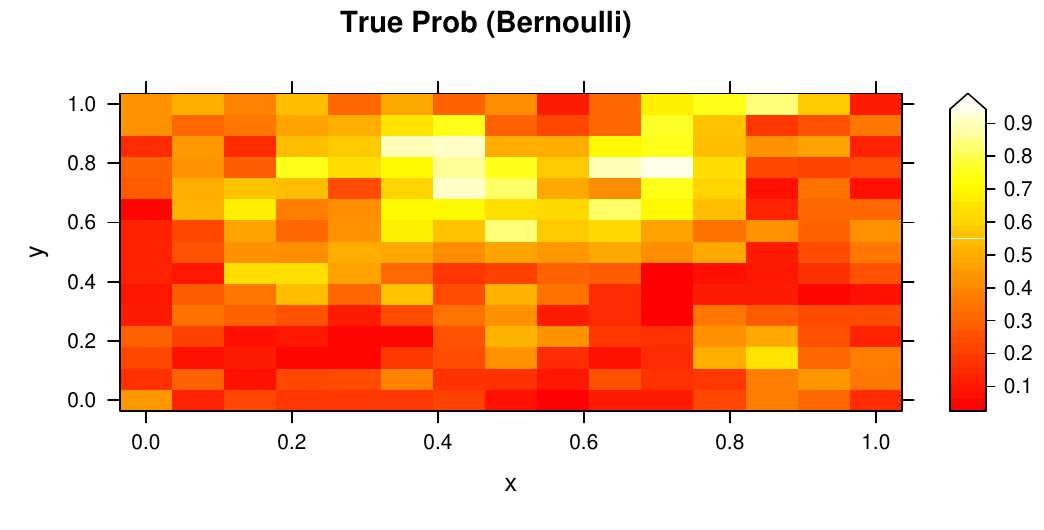}&\includegraphics[width=3.5cm,height=3.5cm]{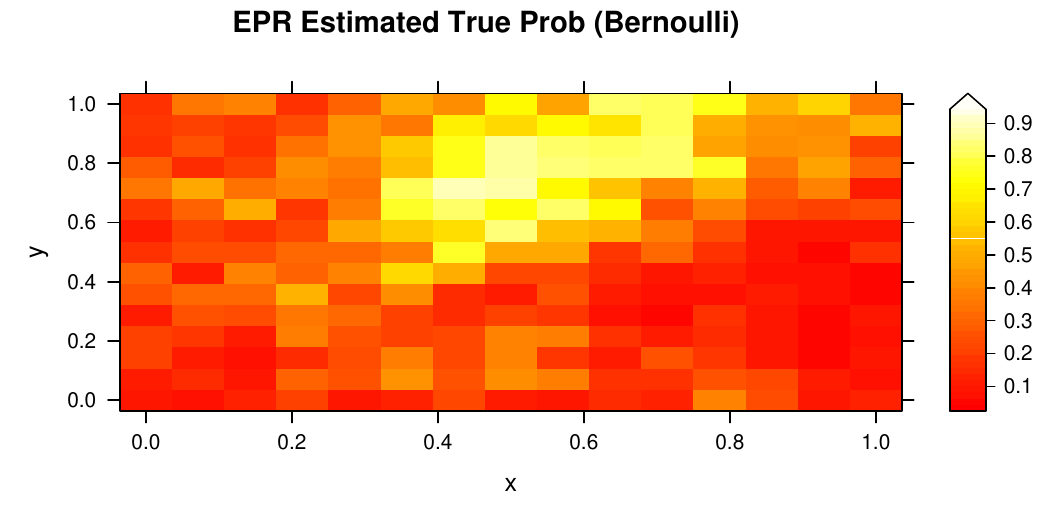}&\includegraphics[width=3.5cm,height=3.5cm]{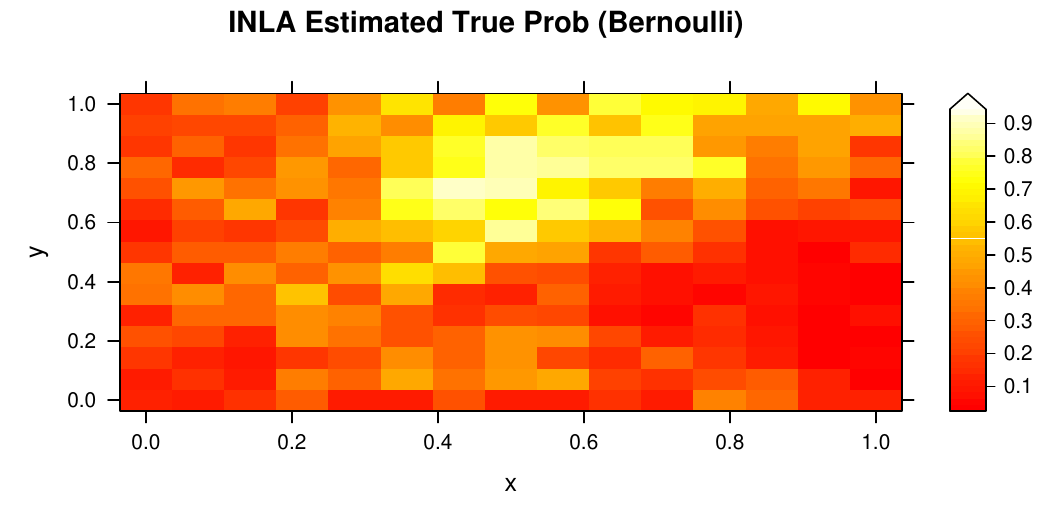}&\includegraphics[width=3.5cm,height=3.5cm]{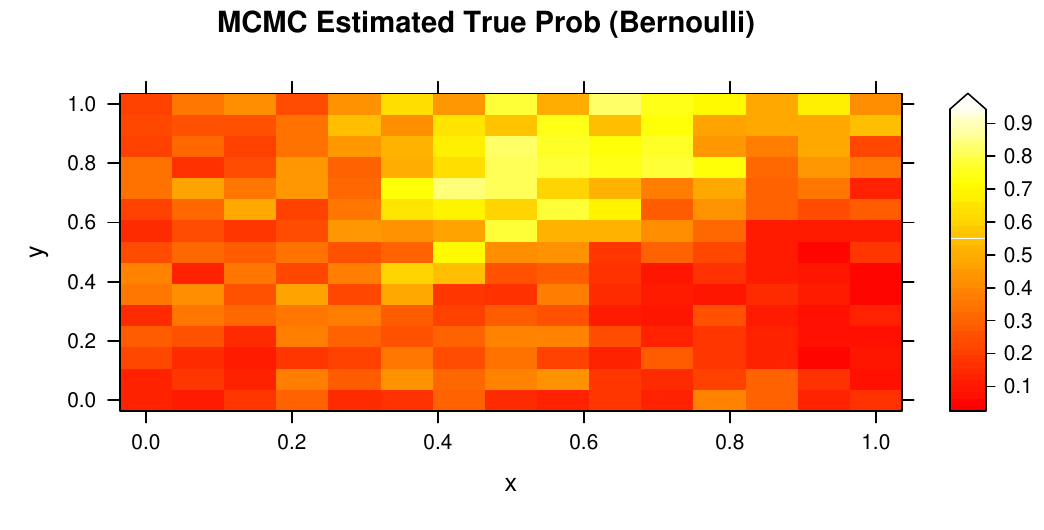}\\
			\includegraphics[width=3.5cm,height=3.5cm]{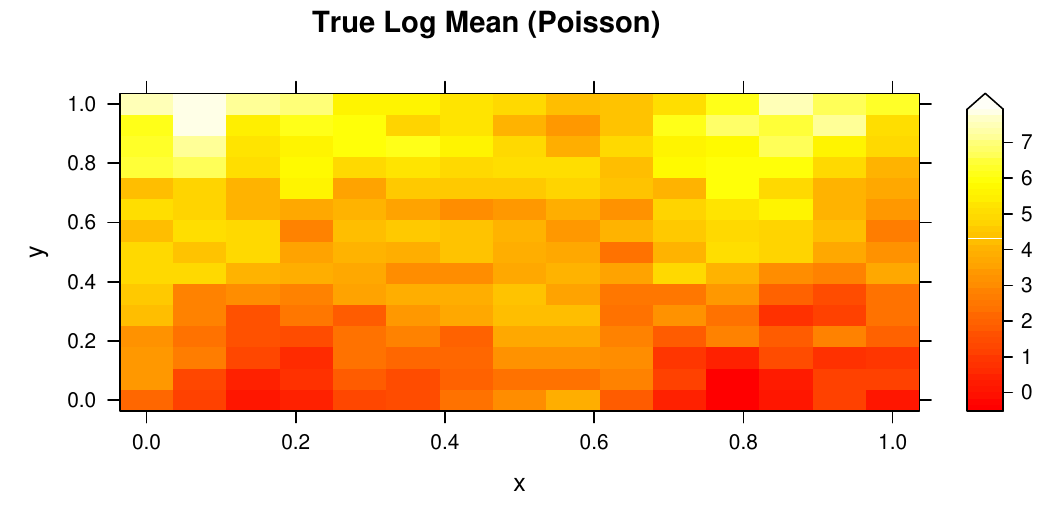}&\includegraphics[width=3.5cm,height=3.5cm]{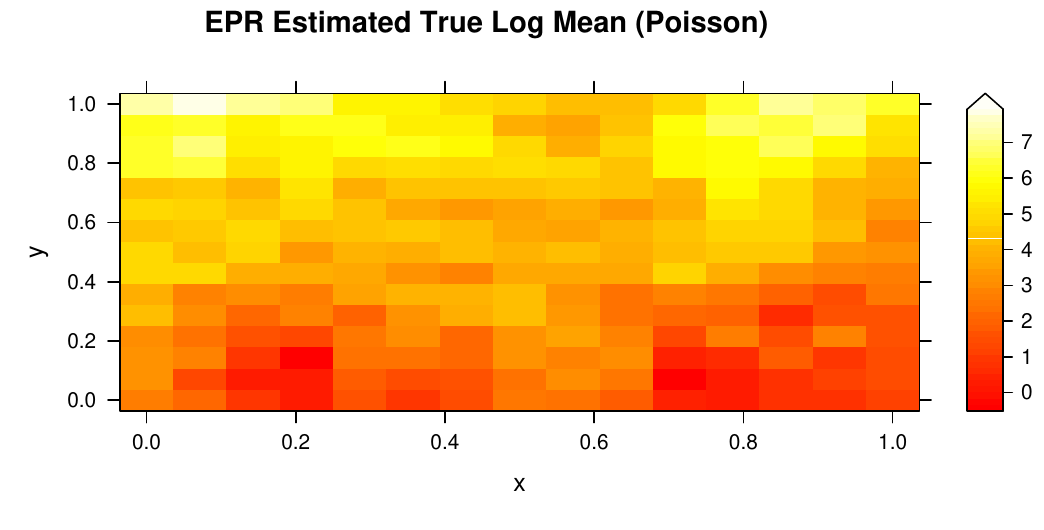}&\includegraphics[width=3.5cm,height=3.5cm]{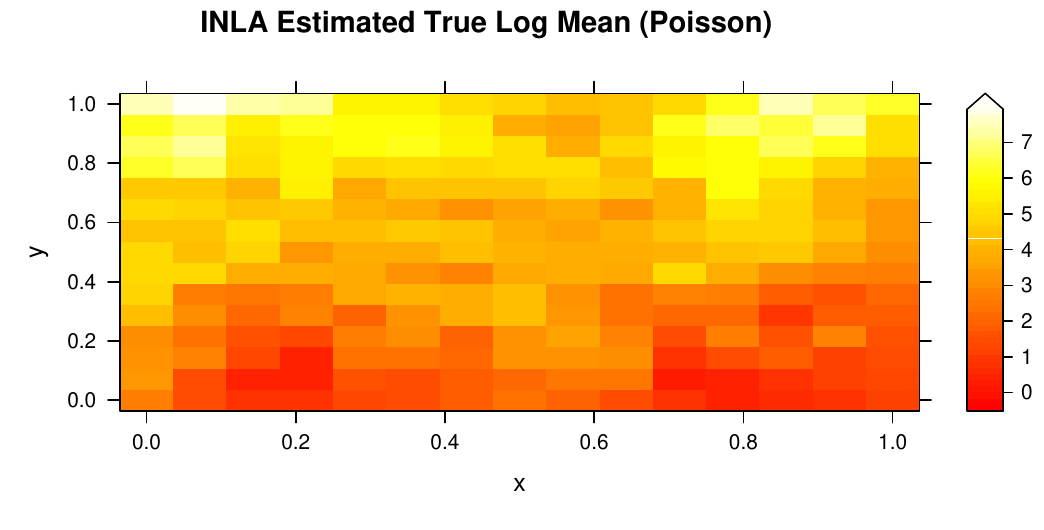}&\includegraphics[width=3.5cm,height=3.5cm]{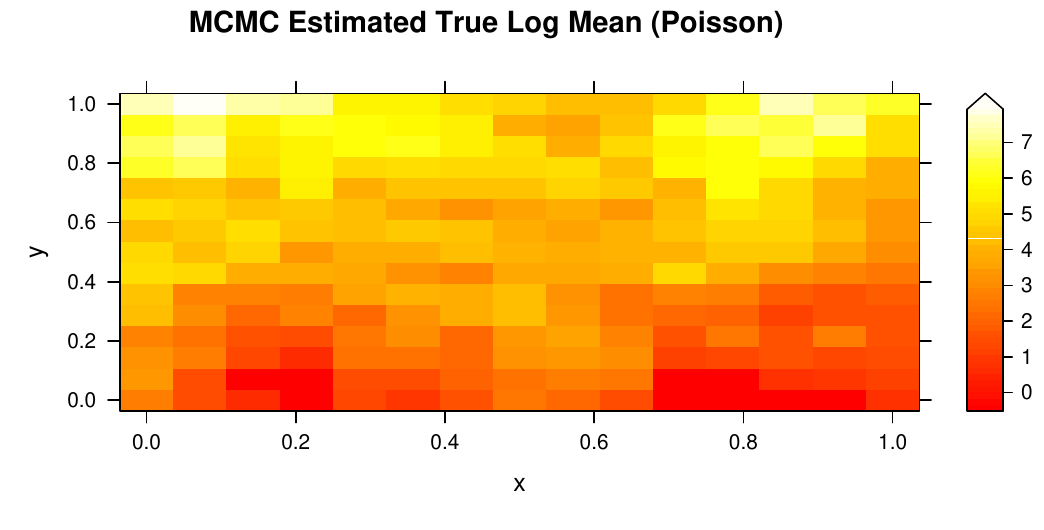}\\
			\includegraphics[width=3.5cm,height=3.5cm]{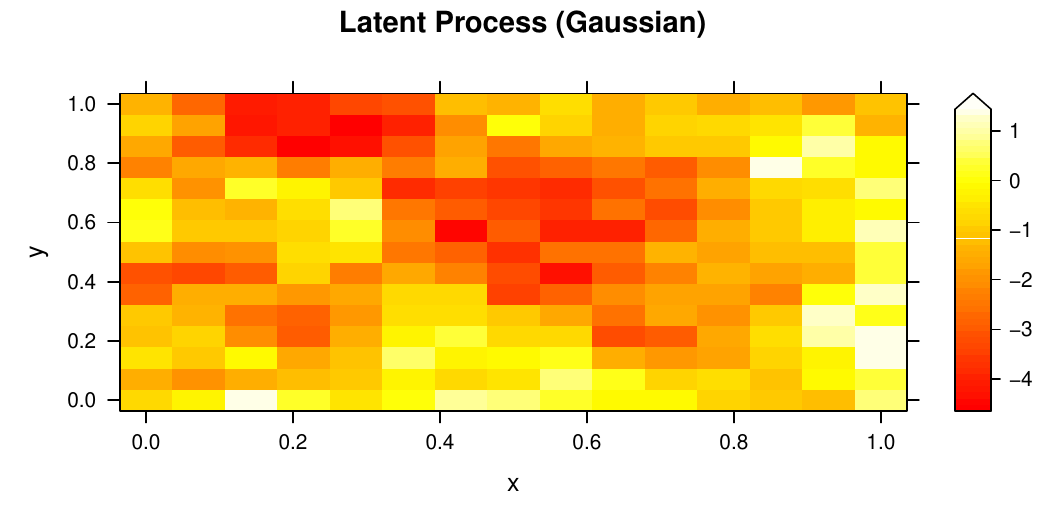}&\includegraphics[width=3.5cm,height=3.5cm]{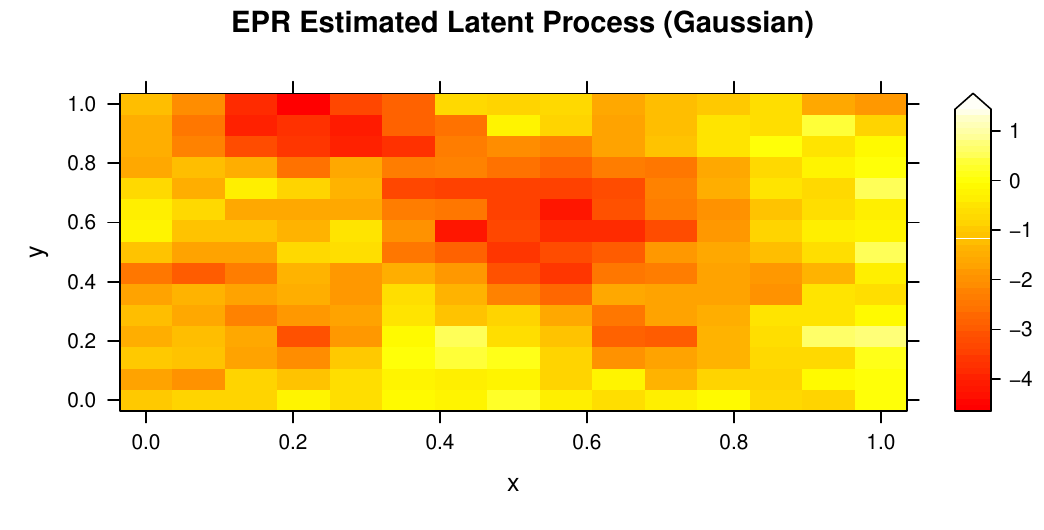}&\includegraphics[width=3.5cm,height=3.5cm]{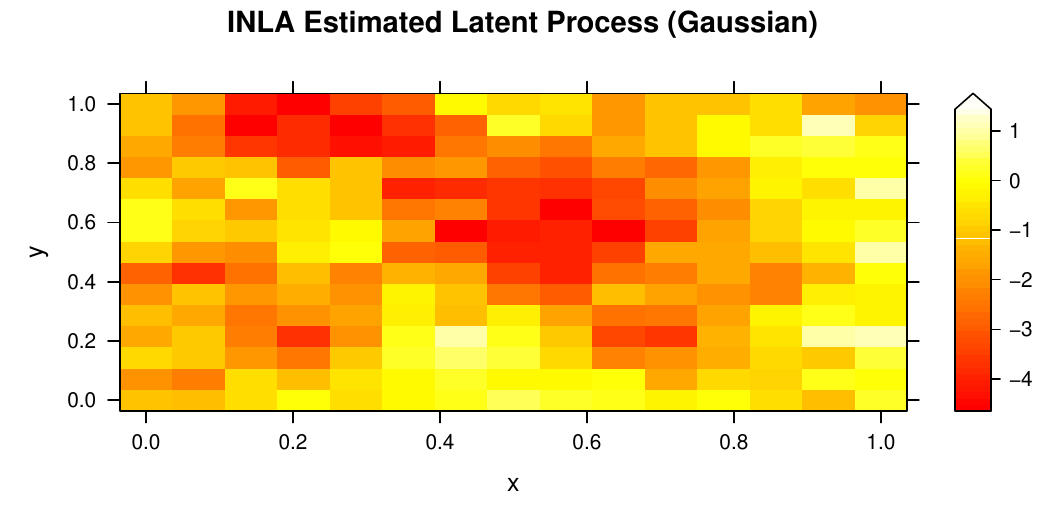}&\includegraphics[width=3.5cm,height=3.5cm]{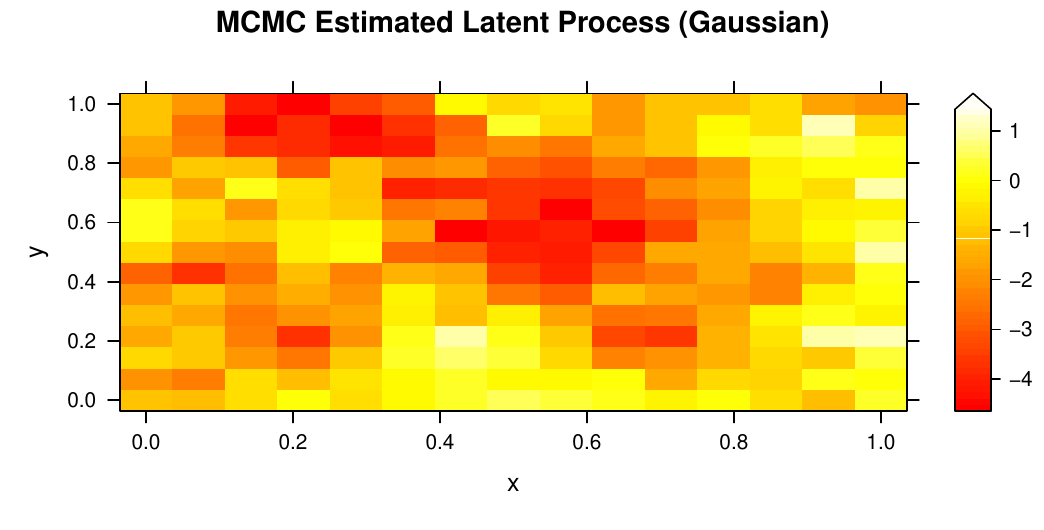}
		\end{tabular}
		
	\end{center}
	\caption{Illustration of EPR, INLA, and MCMC predictions for weakly stationary processes. The first row presents results for binary spatial data, the second row presents results for Poisson spatial data, and the third row presents results for Gaussian spatial data. The left column contains the latent process on the inverse link scale. Second, Third, and Fourth columns display the posterior mean when using EPR, INLA, and MCMC, respectively.}\label{fig:predexamp}
\end{figure}
\subsection{Weakly Stationary Spatial Processes}\label{sec:stationary} A classical assumption for spatially referenced data is that the latent spatial process is weakly stationary. In particular, weakly stationary spatial processes have mean zero and the covariance of the process at any two locations is a positive definite function evaluated at the spatial lag, where this covariance function is referred to as a covariogram. In this section, we compare EPR, INLA, and MCMC implemented using the R package \texttt{spBayes} \citep{spBayes} using the exponential covariogram. The exponential covariogram is a well-known choice, but there are several other choices available \citep[e.g, see][among others]{cressie}. The simulated data are generated as follows,
	\begin{figure}[t!]
	\begin{center}
		\begin{tabular}{ccc}
			\includegraphics[width=4cm,height=4cm]{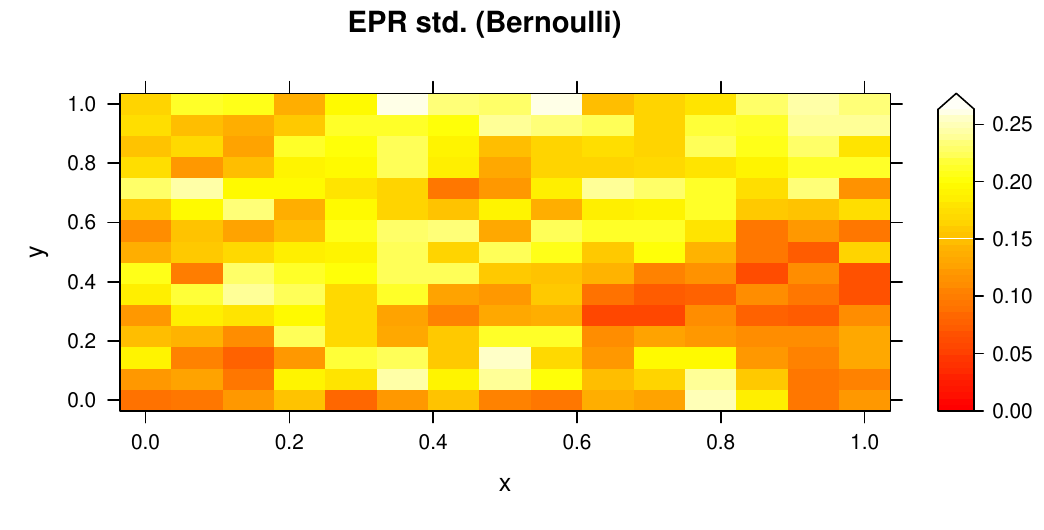}&\includegraphics[width=4cm,height=4cm]{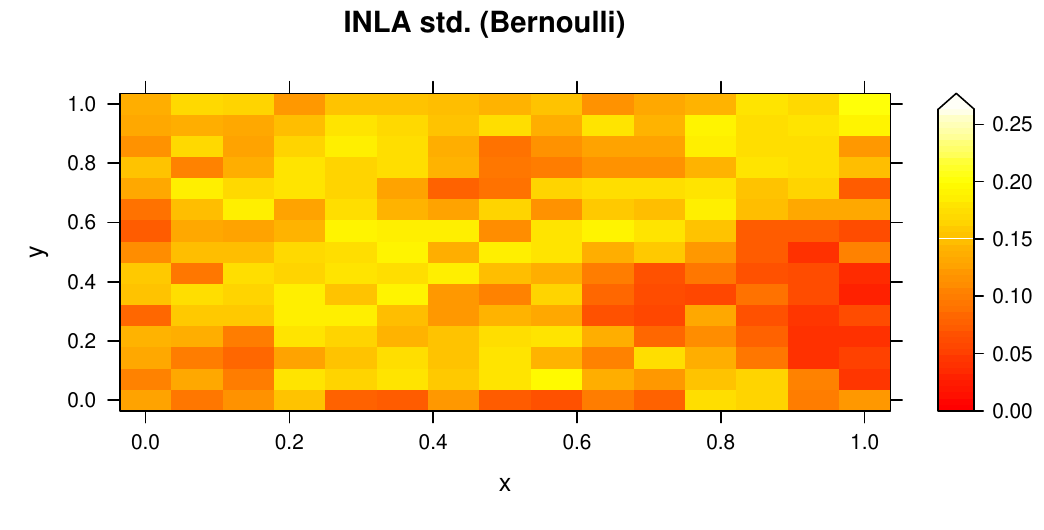}&\includegraphics[width=4cm,height=4cm]{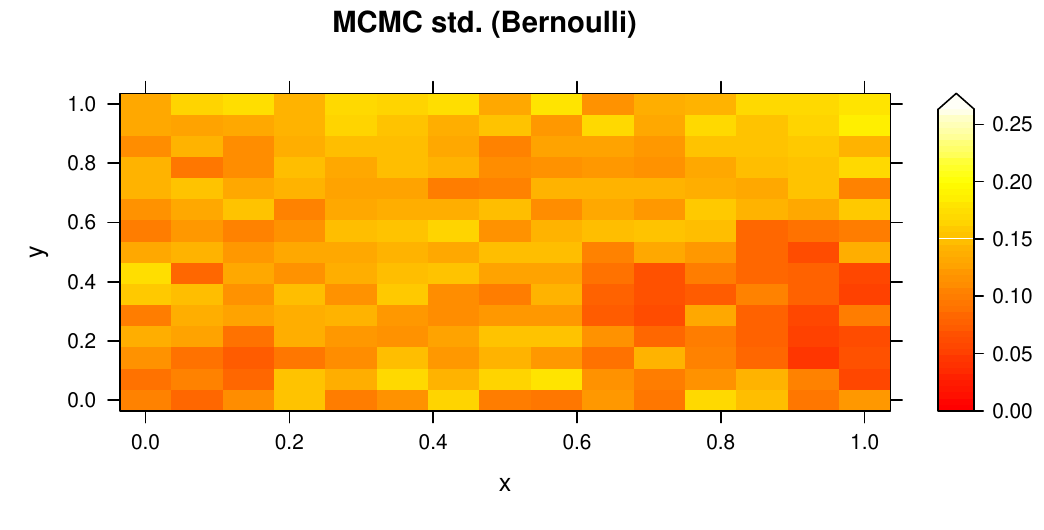}\\
			\includegraphics[width=4cm,height=4cm]{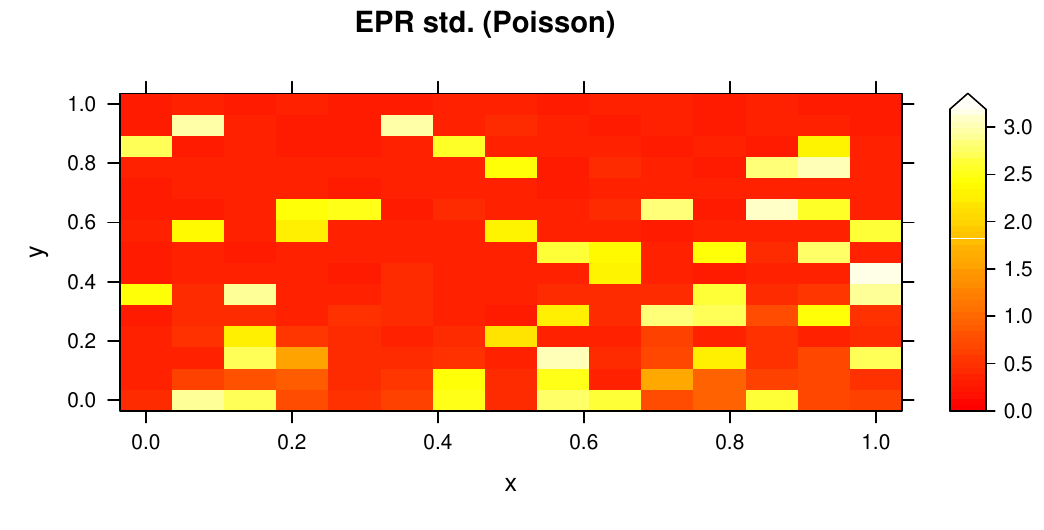}&\includegraphics[width=4cm,height=4cm]{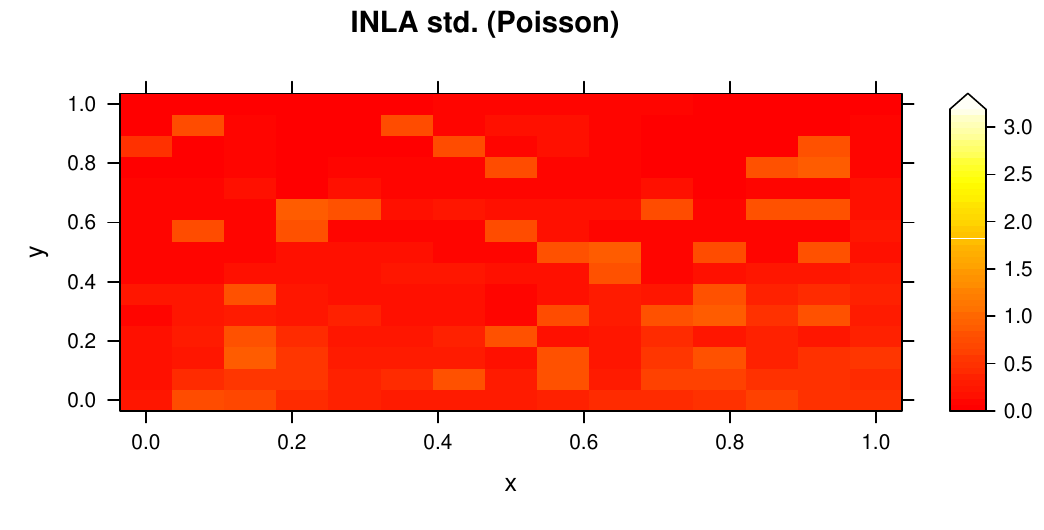}&\includegraphics[width=4cm,height=4cm]{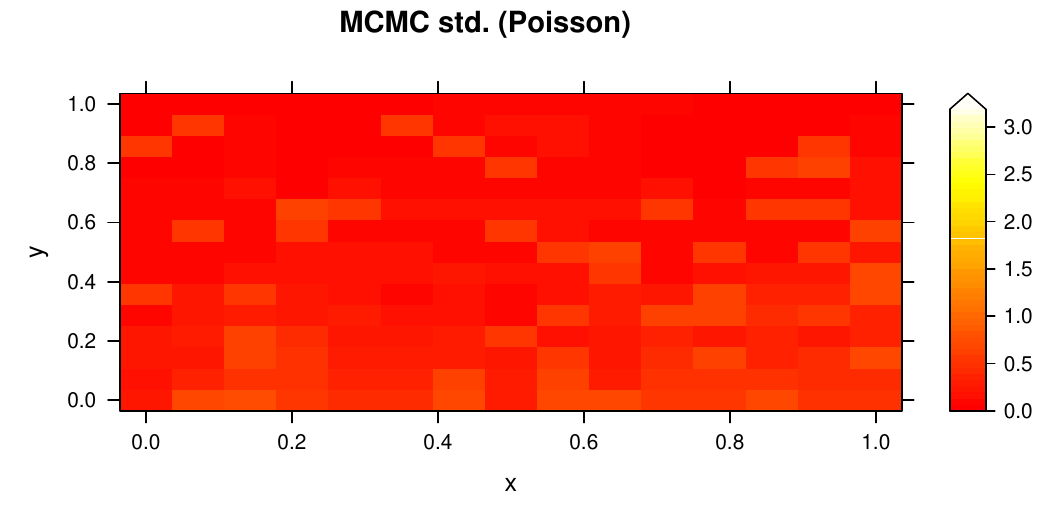}\\
			\includegraphics[width=4cm,height=4cm]{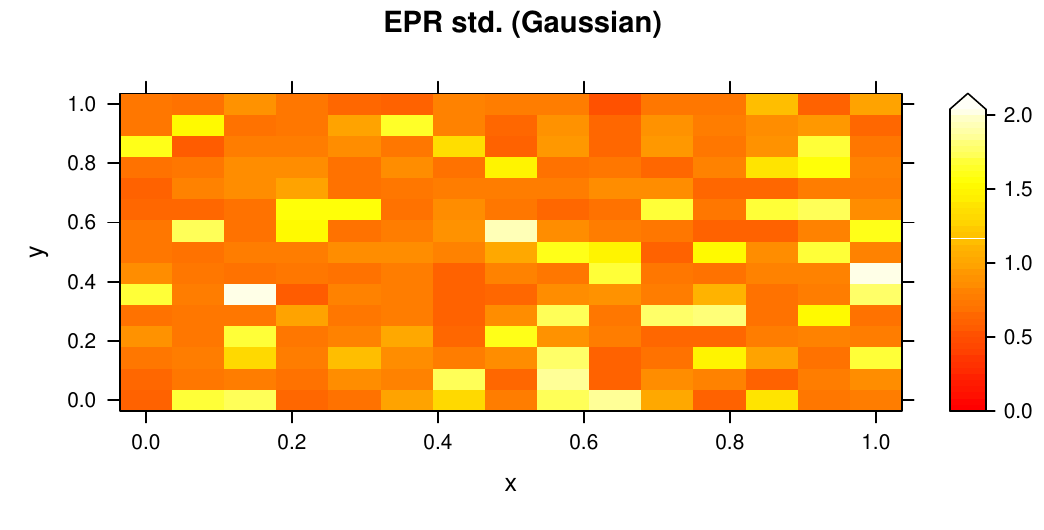}&\includegraphics[width=4cm,height=4cm]{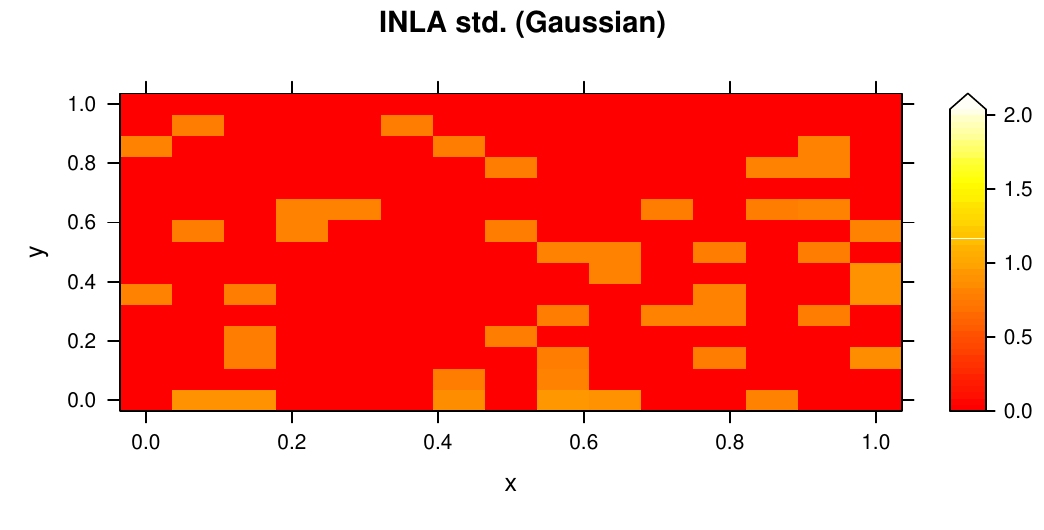}&\includegraphics[width=4cm,height=4cm]{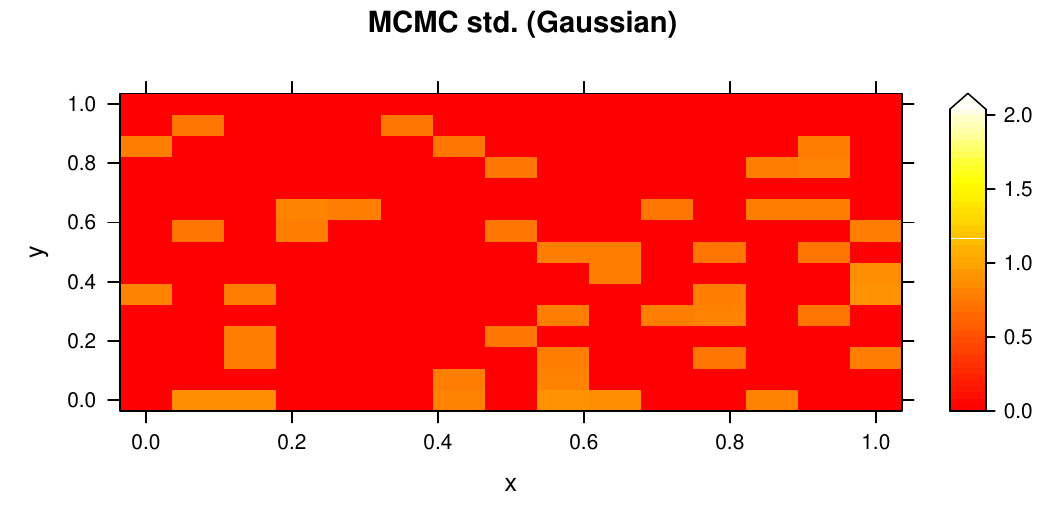}
		\end{tabular}
		
	\end{center}
	\caption{Illustration of EPR, INLA, and MCMC posterior standard deviation for weakly stationary processes for the simulated replicate in the first column of Figure~\ref{fig:predexamp}. The first row presents results for binary spatial data, the second row presents results for Poisson spatial data, and the third row presents results for Gaussian spatial data. The left column contains the latent process on the inverse link scale. Second, Third, and Fourth columns display the posterior mean when using EPR, INLA, and MCMC, respectively.}\label{fig:predvarexamp}
\end{figure}
\begin{align}
\nonumber
Z_{1}(\textbf{s})\vert \nu(\textbf{s})&\sim \mathrm{Normal}\left(-\hspace{2pt}x(\textbf{s}) + \nu(\textbf{s}),0.2\right)\\
\nonumber
Z_{2}(\textbf{s})\vert \nu(\textbf{s})&\sim \mathrm{Poisson}\left\lbrace \mathrm{exp}\left(3+2\hspace{2pt}x(\textbf{s}) + \nu(\textbf{s})\right)\right\rbrace\\
\label{eq:study1}
Z_{3}(\textbf{s})\vert \nu(\textbf{s})&\sim \mathrm{Bernoulli}\left\lbrace \frac{\mathrm{exp}\left(-\hspace{2pt}x(\textbf{s}) + \nu(\textbf{s})\right)}{1+\mathrm{exp}\left(-\hspace{2pt}x(\textbf{s}) + \nu(\textbf{s})\right)}\right\rbrace,
\end{align}
\noindent
where $x(\cdot)$ are generated from a standard uniform distribution of a $15\times 15$ grid of the unit square, and $\nu(\textbf{s})$ is generated as a weakly stationary spatial process with exponential covariogram with range parameter 0.25, nugget variance 0.3, and variance 2 on a $15\times 15$ grid of the unit square. The slope and the intercept for the Poisson example was chosen so that the percent of zero count-valued observations to be small (roughly one percent) to avoid zero inflation. EPR assumes an inverse gamma prior on all variance parameters with shape 1 and rate parameter given a gamma hyperprior with shape and rate set to 1. The range parameter is given a uniform zero to 0.5 prior. The default prior specifications are used for both INLA and spBayes. In Figure~\ref{fig:predexamp}, we provide plots of one simulated replicate and fitted means computed using EPR, INLA, and MCMC. The fitted posterior standard deviations for this example are provided in Figure~\ref{fig:predvarexamp}. In general, we see that all methods perform similarly for this example, however, EPR tends to have larger posterior standard deviation. These patterns are consistent with the example in Section~\ref{sec:sbf}. The fact that the predictions are similar is again notable since, INLA and MCMC are both approximate methods (MCMC is exact in the limit), whereas EPR is an exact MCMC free method.

In Table~\ref{tab:comperform}, we provide the MSPE, MSE, CRPS, and CPU time across computational method and regression type. For logistic regression EPR, INLA, and MCMC perform similarly in terms of MSPE, CRPS, and MSE, since the confidence intervals (CI) over the 50 simulated replicates tend to overlap. In the Poisson data setting, we see similar values of MSPE, MSE, and CRPS, however, there are cases where one approach has a CI that does not overlap. In particular, when comparing CIs for Poisson data, implementation with MCMC leads to a significantly lower MSPE than EPR, implementation with INLA leads to a significantly lower MSE than EPR, and implementation with EPR leads to a significantly lower CRPS than MCMC. In all other cases the CIs overlap in the Poisson data setting. For all three types of spatial linear mixed models MCMC has a considerably larger CPU time than INLA and EPR, and INLA has moderately smaller CPU time than EPR. EPR performs marginally slower than it did in Section~\ref{sec:sbf}, since $\textbf{G}_{M}$ needs to be computed every step of the sampler, whereas the radial basis function in Section~\ref{sec:sbf} only needed to computed once.

\subsection{Intrinsic Conditional Autoregressive Model for American Community Survey Poverty Estimates}\label{sec:car}

The U.S. Census Bureau's ACS provides demographic statistics over several geographies and over 1-year and 5-year time periods \citep{torrieri}. As such, it has become a very useful tool for poverty estimation \citep{molina2010small}. Small area estimation of poverty is a crucial and standard problem in both demography and official statistics \citep{rao2015small}, since it is a key variable for determining economic disparities, public policy, and monitoring the financial circumstances at various levels of geography \citep[e.g., see][]{theil1996studies}. Considering the wide-applicability of EPR, it is important to investigate its performance in standard settings such as poverty estimation. Consequently, in this section, we compare EPR to INLA and MCMC for poverty estimation over U.S. counties in Florida in 2019 using ACS 1-year period estimates. 

Standard demographic related covariates are used; namely, ACS five year period estimates of the median age, the ratio of the population of males to females, and the population (on the log scale) of those who identify as white alone, black or African American alone, and Asian alone. We assume the population of those under the poverty status as binomial distributed with $m_{i}$ representing the $i$-th county's population.  EPR assumes $\sigma_{\xi}^{2} = 0.5$ (chosen with cross validation), and the default prior specifications are used for INLA for the Besag, York, and Molli\'{e} \citep[BYM,][]{besagYorkMollie} model are used. Let $\textbf{A}$ be the row normalized first order binary adjacency matrix. When implementing EPR we define $\textbf{G}$ to be the value such that $\textbf{G}\textbf{G}^{\prime} = \frac{1}{100}(\textbf{I}_{n}-\textbf{A})^{-1}$, which is the Cholesky square root of the covariance matrix implied by the intrinsic conditional autoregressive model with precision $1/100$ (chosen using cross-validation). We fit EPR according to Section~\ref{sec:stepbystep} with $B=100$ independent replicates from the posterior distribution and MCMC using the R package \texttt{CARBayes} with 20,000 replicates with a burn-in of 10,000 \citep{lee2013carbayes}.

\begin{figure}[htp]
	\begin{center}
		\begin{tabular}{ccc}
			\includegraphics[width=6cm,height=6cm]{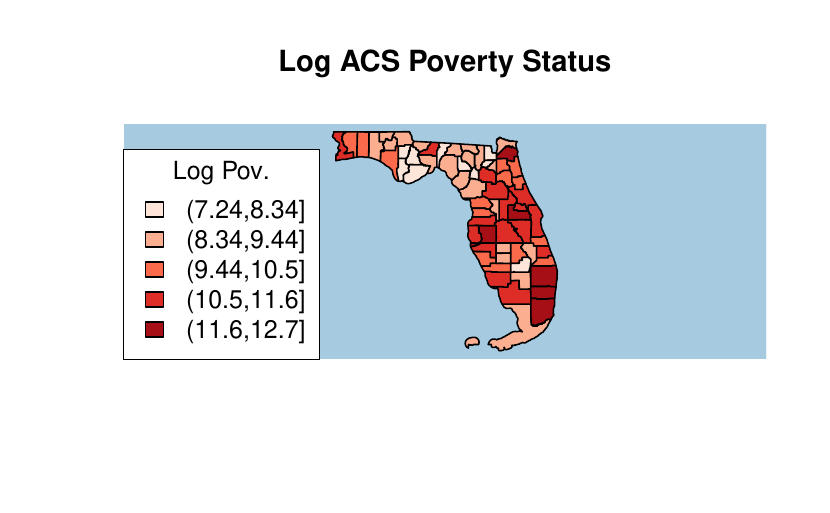}&\hspace{-25pt}\includegraphics[width=6cm,height=6cm]{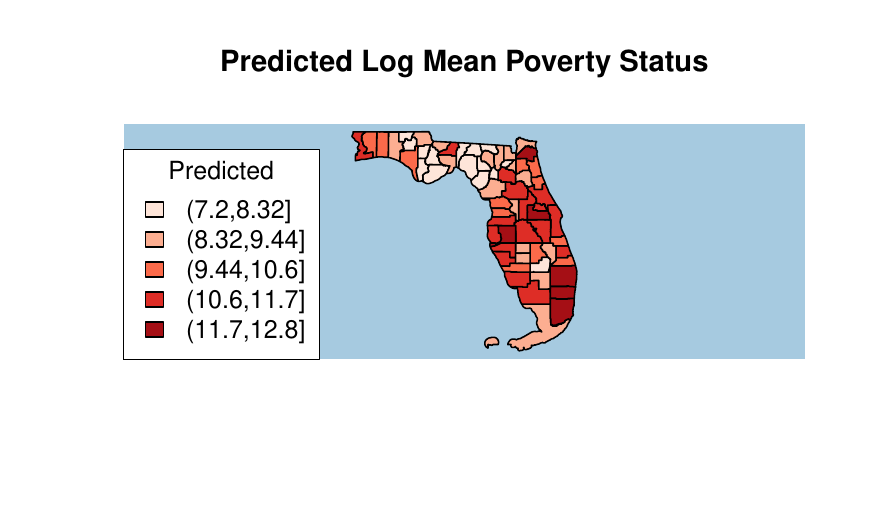}&\hspace{-25pt}
			\includegraphics[width=6cm,height=6cm]{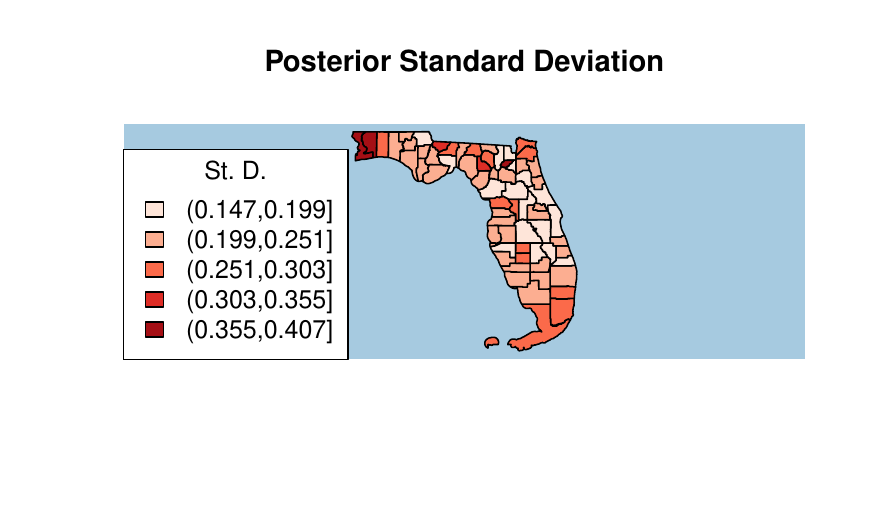}
		\end{tabular}
		\caption{We plot the log ACS estimates (as a reference) along with a plot of the posterior mean and standard deviations computed from 500 independent replicates of EPR. }\label{fig:4}
	\end{center}
\end{figure}

{Plots of the predicted mean and standard deviation of $\widetilde{\textbf{y}}_{k}$ versus the log-data using EPR are provided in Figure~\ref{fig:4}}. In general, we see predictions that reflect the pattern of the data with spatial smoothing. Table~\ref{tab:2} contains several metrics comparing the predictive performance of EPR, INLA, and MCMC. The leave-one-out cross validation error \citep{wahba} is used to assess the predictive performance. Specifically, an observation is left out, and the model is used to predict this value. We compute the relative cross-validation (CV) error and the leave-one-out CRPS. The relative CV suggest that leave-one-out predictions are roughly within 15$\%$ of the hold-out proportion for EPR and INLA, and MCMC has a large relative CV at 70$\%$. These results suggest that EPR and INLA are comparable in terms of CV and CRPS. In general, INLA and EPR produce more similar estimates of the regression coefficients (MSE between these two estimated regression coefficients is 0.1921), and MCMC and INLA produce the most dissimilar regression estimates (MSE between these two estimated regression coefficients is 0.7286). Computationally, EPR is preferable in terms of CPU time followed closely by INLA. MCMC took considerably longer to implement the leave-one-out analysis.

\begin{table}[t!]
	\begin{center}
		\begin{tabular}{@{} l *4c @{}}
			\toprule
			Method  & $\mathrm{CV}$ & CRPS  & CPU\\ 
			\midrule
			EPR & 0.1529 & 0.1668  &18.93 \\ \hline
			INLA &0.1589 & 0.1787 & 83.77  \\\hline
			MCMC & 0.6988 & 0.9858 & 449.65 \\ \bottomrule
		\end{tabular}
	\end{center}
	\caption{ Let $\mathrm{CV}$ be the relative leave-one-out cross-validation error for poverty computed on the logit scale. That is, let $\mathrm{CV}\equiv\underset{i \in \{1,\ldots, n\}}{\mathrm{mean}}\left\lbrace\mathrm{abs}\left(\mathrm{logit}\left(\frac{Z_{i}}{m_{i}}\right) - E_{-i}\left[\widetilde{Y}_{i}\right]\right)/abs(\mathrm{logit}\left(\frac{Z_{i}}{m_{i}}\right))\right\rbrace$, where $E_{-i}$ is the posterior expected value that leaves out $Z_{i}$, ``$abs$'' is the absolute value operator, and ``logit'' is the logit operator. In the column CRPS we evaluate the average CRPS evaluated at the leave one out logit-value. We also provide the CPU time (seconds) to compute the leave-out-out cross-validation criterion.} \label{tab:2}
\end{table}

\section{Discussion} \label{sec:discussion}
This paper describes how to efficiently sample independent replicates directly from the posterior distribution for data modeled using a broad class of spatial latent Gaussian process models. This development required the introduction of the GCM distribution and the conditional GCM distribution. The use of the GCM allows one to consider any class of CM's for their prior distributions on fixed and random effects. Our development explicitly addresses hyperparameters through marginalization. We make use of the GCM in a LGP context to produce what we call exact posterior regression, which represents an efficiently generated independent sample from the posterior distribution. We show that the posterior distribution for fixed and random effects in this LGP are GCM, which we can directly sample from. Furthermore, we use matrix algebra techniques to aid in the computation of EPR.

The results in this paper solve an important problem for Bayesian analysis that is regularly overlooked (i.e., obtaining efficient independent replicates directly from the posterior distribution in Bayesian spatial LGPs). One might also consider empirical Bayesian variations of EPR as our solution also allows one to sample independent replicates directly from the posterior predictive distribution when using point mass specification of $\pi$ (i.e.,point mass on an estimate), which avoids MCMC in empirical Bayesian settings as well. Specifically, Theorems \ref{thm:4} $\--$ \ref{thm:5} can be used with a plug-in estimator of $\bm{\theta}$. However, plug-in estimators have unchecked sampling variability (provided that the plug-in estimator is a non-constant function of the data), and the development of the GCM provides a straightforward solution that accounts for all sources of variability.


While we feel that the results in this manuscript represent a significant advancement in Bayesian modeling of spatial LGPs, it is important to state that MCMC and INLA will always be a standard tool. This is because the spatial LGP specification we consider does not represent the wide variety of LGPs used in the literature (e.g., allowing for mixture components, inference on hyperparameters, data models outside the class of distributions we consider, etc.). Moreover, inference using EPR is limited to summaries of $\bm{\beta}$ and $\widetilde{\textbf{y}}$, since $\bm{\theta}$ is marginalized. However, we hope the theory developed in this article leads to further theoretical developments that allows one to sample independent replicates from the posterior distribution in other settings.

\section*{Acknowledgments}
Jonathan R. Bradley's research was partially supported by the U.S. National Science Foundation
(NSF) under NSF grant SES-1853099. The authors are deeply appreciative to several helpful discussions from Drs. Scott H. Holan and Christopher K. Wikle at the University of Missouri.

\section*{Appendix A: Review of the Conjugate Multivariate Distribution}\label{sec:cm}
In this section, we give the reader a review of the CM distribution from \citet{bradleyLCM}. Suppose the observed data is distributed according to the natural exponential family \citep{diaconis,casellehman}. That is, suppose that the probability density function/probability mass function (pdf/pmf) of the observed datum $Z$ is given by,
\begin{equation}
\label{EF}
f(Z\vert Y,b_{k}) =\mathrm{exp}\left\lbrace ZY - b_{k}\psi_{k}(Y) + c_{k}(Z)\right\rbrace; \hspace{4pt} Z_ \in \mathcal{Z}_{k}, Y \in \mathcal{Y}_{k},
\end{equation}
where $f$ denotes a generic pdf/pmf, $\mathcal{Z}$ is the support of $Z_{k}$, $\mathcal{Y}$ is the support of the unknown parameter $Y$, $b_{k}$ is a possibly known real-value, both $\psi_{k}(\cdot)$ and $c_{k}(\cdot)$ are known real-valued functions, and $k = 1,\ldots, K$ is used to index the specific member of the exponential family (e.g., Gaussian, Poisson, binomial, etc.). The function $b_{k}\psi_{k}(Y)$ is often called the log partition function \citep{casellehman}. We focus on Gaussian responses, which sets $\psi_{1}(Y) = {Y}^{2}$, $b_{1} = 1/2\sigma^{2}$, $\mathcal{Z}_{1} = \mathbb{R}$, and $\mathcal{Y}_{1} = \mathbb{R}$ with $\sigma^{2}>0$;  Poisson responses, which sets $\psi_{2}(Y) = \mathrm{exp}(Y)$, $b_{2} = 1$, $\mathcal{Z}_{2} = \{0,1,2,\ldots\}$, and $\mathcal{Y}_{2} = \mathbb{R}$; and binomial responses, which sets $\psi_{3}(Y) = \mathrm{log}\{1+\mathrm{exp}(Y)\}$, $b_{3} = m$, $\mathcal{Z}_{3} = \{0,1,\ldots, m\}$, and $\mathcal{Y}_{3} = \mathbb{R}$ with $m$ a strictly positive integer.

It follows from \citet{diaconis} that the conjugate prior distribution for $Y$ (when it exists) is given by,
\begin{equation}\label{univ_LG}
f(Y\vert \alpha, \kappa) = \mathcal{N}_{k}(\alpha, \kappa)\hspace{2pt}\mathrm{exp}\left\lbrace \alpha Y - \kappa \psi_{k}(Y)\right\rbrace; \hspace{4pt} Y \in \mathcal{Y}_{k}, \frac{\alpha}{\kappa} \in \mathcal{Z}_{k}, \kappa > 0, k = 1,\ldots, 3,
\end{equation}
\noindent
where $\mathcal{N}_{k}(\alpha, \kappa)$ is a normalizing constant. Let $\mathrm{DY}(\alpha,\kappa;\hspace{2pt}\psi_{k})$ denote a shorthand for the pdf in (\ref{univ_LG}). Here ``DY'' stands for ``Diaconis-Ylvisaker.'' Of course, there are several special cases of the DY distribution other than the Gaussian ($k = 1$), log-gamma ($k = 2$), and logit-beta ($k = 3$) distributions; however, we focus our attention on these standard conjugate cases.

By conjugate we mean that the posterior distribution is from the same family of distributions as the prior distribution. In the case of (\ref{EF}) and (\ref{univ_LG}), we obtain conjugacy as
\begin{align}\label{saturated}
Y\vert Z, \alpha, \kappa  &\sim \mathrm{DY}\left( \alpha + Z, \kappa + b_{k};\hspace{2pt} \psi_{k}\right).
\end{align}
\noindent
\citet{bradleyLCM} derived a multivariate version of $\mathrm{DY}(\alpha,\kappa;\hspace{2pt}\psi_{k})$.  Define the $n$-dimensional random vector $\textbf{y}$ using the following transformation:
\begin{equation}\label{chov}
\textbf{y} = \bm{\mu} + \textbf{V}\textbf{w},
\end{equation}
\noindent
where $\bm{\mu}$ is an $n$-dimensional real-valued vector called the ``location vector,'' $\textbf{V}$ is an $n\times n$ real-valued invertible ``covariance parameter matrix,'' the elements of the $n$-dimensional random vector $\textbf{w}$ are mutually independent, and the $i$-th element of $\textbf{w}$ is $\mathrm{DY}(\alpha,\kappa;\hspace{2pt}\psi_{k})$ with $\alpha_{i}/\kappa_{i}\in \mathcal{Z}_{k}$ and shape/scale (depending on $k$) $\kappa_{i}>0$, respectively. Straightforward change-of-variables of the transformation in (\ref{chov}) yields the following expression for the pdf of $\textbf{y}$:
\begin{align}
\label{mlg_pdf}
& f(\textbf{y}\vert \bm{\mu},\textbf{V},\bm{\alpha},\bm{\kappa}) =\mathrm{det}(\textbf{V}_{k}^{-1})\left\lbrace\prod_{i = 1}^{n}\mathcal{N}_{k}(\kappa_{i},{\alpha_{i}})\right\rbrace\mathrm{exp}\left[\bm{\alpha}^{\prime}\textbf{V}^{-1}(\textbf{y} - \bm{\mu}) - \bm{\kappa}^{\prime}\psi_{k}\left\lbrace\textbf{V}^{-1}(\textbf{y} - \bm{\mu})\right\rbrace\right],
\end{align}
\noindent
where the $j$-th element of $\psi_{k}\left\lbrace\textbf{V}^{-1}(\textbf{y} - \bm{\mu})\right\rbrace$ contains $\psi_{k}$ evaluated at the $j$-th element of the $n$-dimensional vector $\textbf{V}^{-1}(\textbf{y} - \bm{\mu})$, ``det'' denotes the determinant function, $\bm{\alpha}\equiv (\alpha_{1},\ldots,\alpha_{n})^{\prime}$, and $\bm{\kappa} \equiv (\kappa_{1},\ldots,\kappa_{n})^{\prime}$. The density in (\ref{mlg_pdf}) is referred to as the CM distribution, and we use the shorthand $\mathrm{CM}(\bm{\alpha},\bm{\kappa},\bm{\mu},\textbf{V}; {\psi}_{k})$.


\section*{Appendix B : Proofs}

\noindent
{\bf Proof Theorem~\ref{thm:1}}\\
We first derive $f(\textbf{y}\vert \bm{\mu}_{M},\bm{\alpha}_{M},\bm{\kappa}_{M},\textbf{V}_{M},\bm{\theta})$. Upon multiplying independent DY random variables we see that the distribution of $\textbf{w}_{M}$ is,
\begin{equation}
\nonumber
f(\textbf{w}_{M}\vert \bm{\alpha}_{M},\bm{\kappa}_{M}) =\left\lbrace\prod_{k = 1}^{K}\prod_{i = 1}^{n_{k}}\mathcal{N}_{k}(\kappa_{k,i},{\alpha_{k,i}})\right\rbrace\mathrm{exp}\left\lbrace\bm{\alpha}_{M}^{\prime}\textbf{w}_{M} - \bm{\kappa}_{M}^{\prime}\bm{\psi}_{M}\left(\textbf{w}_{M}\right)\right\rbrace,
\end{equation}
The inverse transform is  $\textbf{w}_{M} = \textbf{D}(\bm{\theta})^{-1}\textbf{V}_{M}^{-1}(\textbf{y} - \bm{\mu}_{M})$, and the corresponding Jacobian is $\mathrm{det}\left\lbrace\textbf{D}(\bm{\theta})^{-1}\right\rbrace\mathrm{det}(\textbf{V}_{M}^{-1})$. By standard change-of-variables \citep[e.g., see][]{casellaBerger}, we have that,
\begin{align}
\nonumber
&f(\textbf{y}\vert \bm{\mu}_{M},\textbf{V}_{M},\bm{\alpha}_{M},\bm{\kappa}_{M},\bm{\theta})\\
\nonumber &=\mathrm{det}\left\lbrace\textbf{D}(\bm{\theta})^{-1}\right\rbrace\mathrm{det}(\textbf{V}_{M}^{-1})\left\lbrace\prod_{k = 1}^{K}\prod_{i = 1}^{n_{k}}\mathcal{N}_{k}(\kappa_{k,i},{\alpha_{k,i}})\right\rbrace\\
\label{changofvar}
&\mathrm{exp}\left[\bm{\alpha}_{M}^{\prime}\textbf{D}(\bm{\theta})^{-1}\textbf{V}_{M}^{-1}(\textbf{y} - \bm{\mu}_{M}) - \bm{\kappa}_{M}^{\prime}\bm{\psi}_{M}\left\lbrace\textbf{D}(\bm{\theta})^{-1}\textbf{V}_{M}^{-1}(\textbf{y} - \bm{\mu}_{M})\right\rbrace\right].
\end{align}
\noindent
From our independence assumption, 
\begin{align*} 
f(\textbf{y}\vert \bm{\mu}_{M},\textbf{V}_{M},\bm{\alpha}_{M},\bm{\kappa}_{M}) &= \int_{\Omega} f(\bm{\theta}\vert \bm{\mu}_{M},\textbf{V}_{M},\bm{\alpha}_{M},\bm{\kappa}_{M})f(\textbf{y}\vert \bm{\mu}_{M},\textbf{V}_{M},\bm{\alpha}_{M},\bm{\kappa}_{M},\bm{\theta})d\bm{\theta} \\
&= \int_{\Omega} f(\bm{\theta})f(\textbf{y}\vert \bm{\mu}_{M},\textbf{V}_{M},\bm{\alpha}_{M},\bm{\kappa}_{M},\bm{\theta})d\bm{\theta},
\end{align*}
which upon substituting (\ref{changofvar}) completes the result.\\

\noindent
{\bf Proof Theorem~\ref{thm:2}}\\
From Theorem~2.1, the conditional distribution is given by

\begin{align}
\nonumber
&f(\textbf{y}^{(1)},\bm{\theta}\vert \textbf{y}^{(2)},\bm{\mu}_{M},\textbf{V}_{M},\bm{\alpha}_{M},\bm{\kappa}_{M}) \propto f(\textbf{y}\vert \bm{\mu}_{M},\textbf{V}_{M},\bm{\alpha}_{M},\bm{\kappa}_{M},\bm{\theta})f(\bm{\theta}),\\
\nonumber
&\propto\hspace{5pt}\frac{f(\bm{\theta})}{\mathrm{det}\left\lbrace \textbf{D}(\bm{\theta})\right\rbrace}\mathrm{exp}\left[\bm{\alpha}_{M}^{\prime}\textbf{D}(\bm{\theta})^{-1}\left(\textbf{H}\hspace{6pt} \textbf{Q}\right)  \left\lbrace\left(\begin{matrix}
\textbf{y}^{(1)}\\\textbf{y}^{(2)}
\end{matrix} \right)-\bm{\mu}_{M}\right\rbrace\right.\\
\nonumber
&\hspace{120pt} -\left. \bm{\kappa}_{M}^{\prime}\bm{\psi}_{M}\left\lbrace\textbf{D}(\bm{\theta})^{-1}\left(\textbf{H}\hspace{6pt}\textbf{Q}\right)\left(\begin{matrix}
\textbf{y}^{(1)}\\\textbf{y}^{(2)}
\end{matrix} \right)  - \textbf{D}(\bm{\theta})^{-1}\textbf{V}_{M}^{-1}\bm{\mu}_{M}\right\rbrace\right],\\
\nonumber
&=\hspace{5pt}\frac{f(\bm{\theta})}{\mathrm{det}\left\lbrace \textbf{D}(\bm{\theta})\right\rbrace}\mathrm{exp}\left\lbrace\bm{\alpha}_{M}^{\prime}\textbf{D}(\bm{\theta})^{-1}\textbf{H}\textbf{y}^{(1)}-\bm{\alpha}_{M}^{\prime}\bm{\mu}_{M}^{*} - \bm{\kappa}_{M}^{\prime}\bm{\psi}_{M}\left(\textbf{D}(\bm{\theta})^{-1}\textbf{H}\textbf{y}^{(1)}-\bm{\mu}_{M}^{*}\right)\right\rbrace.
\end{align}
\noindent
Integrating across  $\bm{\theta}$ completes the result.\\

\noindent
{\bf Proof Theorem~\ref{thm:4}}\\
Our strategy is to show that $f(\bm{\zeta},\textbf{q}\vert \textbf{z})\propto \int_{\Omega}\pi(\bm{\theta})f(\bm{\zeta},\textbf{q}, \textbf{z}\vert \bm{\theta})d\bm{\theta}$ is the GCM stated in Theorem~3.1. The data model can be written as:
\begin{equation}
\nonumber
f(\textbf{z}\vert \bm{\xi},\bm{\beta}, \bm{\eta},\{\sigma_{i}\}, \bm{\delta}_{y}) \propto {N}\hspace{2pt}\mathrm{exp}\left[\textbf{a}^{\prime}\left(\textbf{I}_{n}, \textbf{X}, \textbf{G}\right) \left\lbrace\bm{\zeta}-\bm{\delta}_{y}\right\rbrace-\textbf{b}^{\prime}\bm{\psi}_{D}\left\lbrace \left(\textbf{I}_{n}, \textbf{X}, \textbf{G}\right) \bm{\zeta} - \bm{\delta}_{y}\right\rbrace\right],
\end{equation}
\noindent
where ${N} = \prod_{i = 1}^{n}\frac{1}{\sigma_{i}}$ for the Gaussian setting, and ${N} = 1$ in the binomial and Poisson setting. Recall $\bm{\zeta} = (\bm{\xi}^{\prime}, \bm{\beta}^{\prime}, \bm{\eta}^{\prime})^{\prime}$, and $\textbf{a} = \textbf{D}_{\sigma}^{(\prime)}\textbf{z}$ when $\textbf{z}$ is Gaussian distributed, $\textbf{a} =\textbf{z}$ when $\textbf{z}$ is Poisson or binomial distributed, $\psi_{D} (\cdot) = \psi_{1}(\cdot)\bm{1}_{1,n}\textbf{D}_{\sigma}$ when $\textbf{z}$ is Gaussian distributed, and $\psi_{D}(\cdot) = \psi_{k}(\cdot)\bm{1}_{1,n}$ when $\textbf{z}$ is Poisson or binomial distributed.  The density $f(\bm{\zeta},\textbf{q},\bm{\theta}\vert \textbf{z}, \bm{\alpha}_{\beta},\bm{\alpha}_{\eta},\bm{\kappa}_{\beta},\bm{\kappa}_{\eta})$ is proportional to the product 
\begin{equation}
\nonumber
f(\textbf{z}\vert \bm{\xi},\bm{\beta}, \bm{\eta}, \bm{\delta}_{y})f(\bm{\xi}\vert \bm{\beta},\bm{\eta},\bm{\delta}_{y},\bm{\delta}_{\xi})f(\bm{\beta}\vert  \bm{\alpha}_{\beta},\bm{\kappa}_{\beta},\bm{\delta}_{\beta},\textbf{D}_{\beta}(\bm{\theta}))f(\bm{\eta}\vert \bm{\alpha}_{\eta},\bm{\kappa}_{\eta},\bm{\delta}_{\eta},\textbf{D}_{\eta}(\bm{\theta}))f(\textbf{q})\pi(\bm{\theta}),
\end{equation}
\noindent
where recall $\bm{\delta} = (\bm{\delta}_{y}^{\prime},\bm{\delta}_{\beta}^{\prime},\bm{\delta}_{\eta}^{\prime},\bm{\delta}_{\xi}^{\prime})^{\prime} = -\textbf{D}(\bm{\theta})^{-1}\textbf{Q}\textbf{q}$. Now,
\begin{align}
\nonumber
&f(\bm{\xi}\vert \bm{\beta},\bm{\eta},\bm{\delta}_{y},\bm{\delta}_{\xi})\propto\\
\nonumber
& \hspace{20pt}\mathrm{exp}\left[\bm{\alpha}_{\xi}^{\prime}\left\lbrace\left(\begin{array}{ccc}
\textbf{I}_{n} & \textbf{X} & \textbf{G}\\
\frac{1}{\sigma_{\xi}^{2}}\textbf{I}_{n} &  \bm{0}_{n,p} & \bm{0}_{n,r}
\end{array}
\right) \bm{\zeta}-\left(\begin{array}{c}
\bm{\delta}_{y}\\
\bm{\delta}_{\xi}
\end{array}\right)\right\rbrace\right.\\
\nonumber
&\hspace{80pt}\left.-\bm{\kappa}_{\xi}^{\prime}\bm{\psi}_{D,\xi}\left\lbrace \left(\begin{array}{ccc}
\textbf{I}_{n} & \textbf{X} & \textbf{G}\\
\frac{1}{\sigma_{\xi}^{2}}\textbf{I}_{n} &  \bm{0}_{n,p} & \bm{0}_{n,r}
\end{array}
\right) \bm{\zeta}-\left(\begin{array}{c}
\bm{\delta}_{y}\\
\bm{\delta}_{\xi}
\end{array}\right)\right\rbrace\right],
\end{align}
\noindent
with $\sigma_{\xi}^{2}$ and $\alpha_{\xi}$ known, $\bm{\alpha}_{\xi} = \bm{0}_{2n,1}$ when $\textbf{z}$ is Gaussian distributed, $\bm{\alpha}_{\xi} = (\bm{1}_{1,n},\bm{0}_{1,n})^{\prime}$ when $\textbf{z}$ is Poisson or binomial distributed, $\bm{\kappa}_{\xi} = (\bm{0}_{1,n},\frac{1}{2}\bm{1}_{1,n})^{\prime}$ when $\textbf{z}$ is Gaussian or Poisson distributed, and $\bm{\kappa}_{\xi} = (2\alpha_{\xi}\bm{1}_{1,n},\frac{1}{2}\bm{1}_{1,n})^{\prime}$ when $\textbf{z}$ is binomial distributed. Let $\bm{\psi}_{D,\xi}(\textbf{h}_{D,\xi}) =\left(\psi_{D}(\textbf{h})^{\prime},\psi_{1}(\textbf{h}^{*})^{\prime}\right)^{\prime}$
for $2n$-dimensional real-valued vector $\textbf{h}_{D,\xi}=(\textbf{h}^{\prime}, \textbf{h}^{*\prime})^{\prime}$. The product,
\begin{align}
\nonumber
&f(\textbf{z}\vert \bm{\xi},\bm{\beta}, \bm{\eta}, \bm{\delta}_{y})f(\bm{\xi}\vert \bm{\beta},\bm{\eta},\bm{\delta}_{y},\bm{\delta}_{\xi})\propto\\
\nonumber
&\hspace{20pt}{N}\hspace{2pt}\mathrm{exp}\left[\textbf{a}_{Z}^{\prime}\left\lbrace\left(\begin{array}{ccc}
\textbf{I}_{n} & \textbf{X} & \textbf{G}\\
\frac{1}{\sigma_{\xi}^{2}}\textbf{I}_{n} &  \bm{0}_{n,p} & \bm{0}_{n,r}
\end{array}
\right) \bm{\zeta}-\left(\begin{array}{c}
\bm{\delta}_{y}\\
\bm{\delta}_{\xi}
\end{array}\right)\right\rbrace\right.\\
\nonumber
&\left. \hspace{80pt}-\textbf{b}_{Z}^{\prime}\bm{\psi}_{D,\xi}\left\lbrace \left(\begin{array}{ccc}
\textbf{I}_{n} & \textbf{X} & \textbf{G}\\
\frac{1}{\sigma_{\xi}^{2}}\textbf{I}_{n} &  \bm{0}_{n,p} & \bm{0}_{n,r}
\end{array}
\right) \bm{\zeta}-\left(\begin{array}{c}
\bm{\delta}_{y}\\
\bm{\delta}_{\xi}
\end{array}\right)\right\rbrace\right],
\end{align}
where $\textbf{a}_{Z} = (\textbf{z}^{\prime}\textbf{D}_{\sigma},\bm{0}_{1,n})^{\prime}$ when $\textbf{z}$ is Gaussian distributed, $\textbf{a}_{Z} = (\textbf{z}^{\prime}+\alpha_{\xi}\bm{1}_{1,n},\bm{0}_{1,n})^{\prime}$ when $\textbf{z}$ is Poisson or binomial distributed, $\textbf{b}_{Z} = (\frac{1}{2}\bm{1}_{1,n}\textbf{D}_{\sigma},\frac{1}{2}\bm{1}_{1,n})^{\prime}$ when $\textbf{z}$ is Gaussian distributed, $\textbf{b}_{Z} = (\bm{1}_{1,n},\frac{1}{2}\bm{1}_{1,n})^{\prime}$ when $\textbf{z}$ is Poisson distributed, and $\textbf{b}_{Z} = (\textbf{m}^{\prime}+2\alpha_{\xi}\bm{1}_{1,n},\frac{1}{2}\bm{1}_{1,n})^{\prime}$ when $\textbf{z}$ is binomial distributed.

Notice that the implied shape/scale parameters $\textbf{a}_{Z}$ and $\textbf{b}_{Z}$ are not on the boundary of the parameter space (when zero counts are present), which is a motivation for including $\bm{\xi}$ in the LGP. Multiplying by $f(\bm{\beta}\vert \bm{\alpha}_{\beta},\bm{\kappa}_{\beta},\bm{\delta}_{\beta},\textbf{D}_{\beta}(\bm{\theta}))f(\bm{\eta}\vert  \bm{\alpha}_{\eta},\bm{\kappa}_{\eta},\bm{\delta}_{\eta},\textbf{D}_{\eta}(\bm{\theta}))\pi(\bm{\theta})$, and stacking vector and matrices leads to 
\begin{align}
\nonumber
& f(\bm{\zeta},\textbf{q},\bm{\theta}\vert \textbf{z}) \propto \\
\nonumber
& \frac{\pi_{*}(\bm{\theta})}{\mathrm{det}\left\lbrace\textbf{D}(\bm{\theta})\right\rbrace}\mathrm{exp}\left[\bm{\alpha}_{M}^{\prime}\left\lbrace\left(\begin{array}{ccc}
\textbf{I}_{n} & \textbf{X} & \textbf{G}\\
\bm{0}_{p,n} & \textbf{D}_{\beta}(\bm{\theta})^{-1} & \bm{0}_{p,r}\\
\bm{0}_{r,n} &  \bm{0}_{r,p} & \textbf{D}_{\eta}(\bm{\theta})^{-1}\\
\frac{1}{\sigma_{\xi}^{2}}\textbf{I}_{n} &  \bm{0}_{n,p} & \bm{0}_{n,r}
\end{array}
\right) \bm{\zeta}-\left(\begin{array}{c}
\bm{\delta}_{y}\\
\bm{\delta}_{\beta}\\
\bm{\delta}_{\eta}\\
\bm{\delta}_{\xi}
\end{array}\right)\right\rbrace \right.\\
\nonumber
&\hspace{80pt}\left. -\bm{\kappa}_{M}^{\prime}\bm{\psi}_{M}\left\lbrace \left(\begin{array}{ccc}
\textbf{I}_{n} & \textbf{X} & \textbf{G}\\
\bm{0}_{p,n} & \textbf{D}_{\beta}(\bm{\theta})^{-1} & \bm{0}_{p,r}\\
\bm{0}_{r,n} &  \bm{0}_{r,p} & \textbf{D}_{\eta}(\bm{\theta})^{-1}\\
\frac{1}{\sigma_{\xi}^{2}}\textbf{I}_{n} &  \bm{0}_{n,p} & \bm{0}_{n,r}
\end{array}
\right) \bm{\zeta}-\left(\begin{array}{c}
\bm{\delta}_{y}\\
\bm{\delta}_{\beta}\\
\bm{\delta}_{\eta}\\
\bm{\delta}_{\xi}
\end{array}\right)\right\rbrace\right].
\end{align} 
Substituting $\bm{\delta} = -\textbf{D}(\bm{\theta})^{-1}\textbf{Q}\textbf{q}$ and integrating with respect to $\bm{\theta}$ leads to
\begin{align}
\nonumber
& f(\bm{\zeta},\textbf{q}\vert \textbf{z}) \propto \\
\nonumber
& \int_{\Omega}\frac{\pi_{*}(\bm{\theta})}{\mathrm{det}\left\lbrace\textbf{D}(\bm{\theta})\right\rbrace}\mathrm{exp}\left[\bm{\alpha}_{M}^{\prime}\textbf{D}(\bm{\theta})^{-1}(\textbf{H},\textbf{Q})\left(\begin{array}{c}
\bm{\zeta}\\
\textbf{q}
\end{array}
\right)-\bm{\kappa}_{M}^{\prime}\bm{\psi}_{M}\left\lbrace \textbf{D}(\bm{\theta})^{-1}(\textbf{H},\textbf{Q})\left(\begin{array}{c}
\bm{\zeta}\\
\textbf{q}
\end{array}
\right)\right\rbrace\right]d\bm{\theta}\\
\nonumber
&\propto \mathrm{GCM}(\bm{\alpha}_{M},\bm{\kappa}_{M}, \bm{0}_{2n+p+r,1},\textbf{V}_{M},\pi_{*},\textbf{D}; \bm{\psi}_{M}),
\end{align} 
\noindent
which completes the result.\\

\noindent
{\bf Proof Theorem~\ref{cor:2}}\\
Equations (\ref{zetasim}) and \ref{qsim}) from the main text follows from (\ref{cholV2}) from the main text, Theorem~3.1, and that
\begin{align*} (\textbf{H},\textbf{Q})^{-1}\textbf{D}(\bm{\theta}) = \left\lbrace \textbf{H}(\textbf{H}^{\prime}\textbf{H})^{-1}, \textbf{Q}\right\rbrace^{\prime}\textbf{D}(\bm{\theta}).
\end{align*}
Let $\textbf{w}_{M}$ consist of independent DY random variables with respective shape and scale parameters in $\bm{\alpha}_{M}$ and $\bm{\kappa}_{M}$. Let
\begin{equation}
\nonumber
\textbf{w} =\textbf{D}(\bm{\theta})\textbf{w}_{M} ,
\end{equation}
where recall $\textbf{D}(\bm{\theta})$ is a block diagonal matrix with first $n\times n$ block diagonal equaling the identity matrix. It follows from Theorem~2.1 that $\textbf{w}$ has the stated GCM distribution in Theorem~3.2. Then
\begin{align*}
&\left(\begin{array}{c}
\bm{\zeta}_{rep}\\
\textbf{q}_{rep}
\end{array}\right)=\left(\begin{array}{c}(\textbf{H}^{\prime}\textbf{H})^{-1}\textbf{H}^{\prime}\\
\textbf{Q}^{\prime}
\end{array}\right)\textbf{w} =  \left(\begin{array}{c}(\textbf{H}^{\prime}\textbf{H})^{-1}\textbf{H}^{\prime}\\
\textbf{Q}^{\prime}
\end{array}\right)\textbf{D}(\bm{\theta})\textbf{w}_{M}.
\end{align*}
From Theorem~2.1 it follows that $(\bm{\zeta}_{rep}^{\prime},
\textbf{q}_{rep}^{\prime})^{\prime}$ is the  GCM stated in Theorem~3.2. Equation (\ref{ysim}) in the main text follows from the fact that $\textbf{y} = (\textbf{I}_{n},\bm{0}_{n,n+p+r})\textbf{H}\bm{\zeta} +  (\textbf{I}_{n},\bm{0}_{n,n+p+r})\textbf{Q}\textbf{q}$ so that
\begin{align}
\nonumber
\textbf{y}_{rep} &= (\textbf{I}_{n},\bm{0}_{n,n+p+r})\textbf{H}\bm{\zeta}_{rep} +  (\textbf{I}_{n},\bm{0}_{n,n+p+r})\textbf{Q}\textbf{q}_{rep} \\
\nonumber
&= (\textbf{I}_{n},\bm{0}_{n,n+p+r})\textbf{H}(\textbf{H}^{\prime}\textbf{H})^{-1}\textbf{H}^{\prime}\textbf{w} +  (\textbf{I}_{n},\bm{0}_{n,n+p+r})\textbf{Q}\textbf{Q}^{\prime}\textbf{w}\\
\nonumber
&=(\textbf{I}_{n},\bm{0}_{n,n+p+r})\textbf{H}(\textbf{H}^{\prime}\textbf{H})^{-1}\textbf{H}^{\prime}\textbf{w} +  (\textbf{I}_{n},\bm{0}_{n,n+p+r})(\textbf{I}-\textbf{H}(\textbf{H}^{\prime}\textbf{H})^{-1}\textbf{H}^{\prime})\textbf{w}\\
\nonumber
&=(\textbf{I}_{n},\bm{0}_{n,n+p+r})\textbf{w} = (\textbf{I}_{n},\bm{0}_{n,n+p+r})\textbf{w} .
\end{align}

%

\noindent
{\bf Proof Theorem~\ref{thm:5}}\\
From \citep{lu2002inverses} the inverse of a 2$\times$2 block matrix is,
\begin{align*}
&\left(\begin{array}{cc}
\textbf{A}_{11} & \textbf{A}_{12}\\
\textbf{A}_{21} & \textbf{A}_{22}
\end{array}
\right)^{-1} =\\
&\left(\begin{array}{cc}
\textbf{A}_{11}^{-1} + \textbf{A}_{11}^{-1}\textbf{A}_{12}(\textbf{A}_{22} - \textbf{A}_{21}\textbf{A}_{11}^{-1}\textbf{A}_{12})^{-1}\textbf{A}_{21}\textbf{A}_{11}^{-1} & -\textbf{A}_{11}^{-1}\textbf{A}_{12}(\textbf{A}_{22} - \textbf{A}_{21}\textbf{A}_{11}^{-1}\textbf{A}_{12})^{-1}\\
-(\textbf{A}_{22} - \textbf{A}_{21}\textbf{A}_{11}^{-1}\textbf{A}_{12})^{-1}\textbf{A}_{21}\textbf{A}_{11}^{-1} & (\textbf{A}_{22} - \textbf{A}_{21}\textbf{A}_{11}^{-1}\textbf{A}_{12})^{-1}
\end{array}
\right),
\end{align*}
\noindent
for generic real-valued $M\times M$ matrix $\textbf{A}_{11}$, $M\times (p+r)$ matrix $\textbf{A}_{12}$, $(p+r)\times M$ matrix $\textbf{A}_{21}$, and $(p+r)\times (p+r)$ matrix $\textbf{A}_{22}$.
Equation (13) of the main text follows from applying this known inverse of 2$\times$2 block matrices\citep{lu2002inverses} to,
\begin{equation}
\nonumber
(\textbf{H}^{\prime}\textbf{H}) = \left(\begin{array}{ccc}
2\textbf{I}_{n}& \textbf{X} & \textbf{G}\\
\textbf{X}^{\prime} & \textbf{X}^{\prime}\textbf{X}+\textbf{I}_{p} & \textbf{X}^{\prime}\textbf{G}\\
\textbf{G}^{\prime} &  \textbf{G}^{\prime}\textbf{X} & \textbf{G}^{\prime}\textbf{G}+\textbf{I}_{r}\\
\end{array}
\right).
\end{equation}
\noindent
Similarly, Equation (14) from the main text follows from applying the same inverse identity to
\begin{equation}
\nonumber
(\textbf{D}-\textbf{B}^{\prime}\textbf{A}^{-1}\textbf{B}) = \left(\begin{array}{cc}
\textbf{A}^{*}& \textbf{B}^{*}\\
\textbf{C}^{*} & \textbf{D}^{*}
\end{array}
\right).
\end{equation}

\noindent
{\bf Proof Theorem~\ref{thm:7}}\\ \citep{lu2002inverses} gave two identities for the inverse of a 2$\times$2 block matrix,
\begin{equation}
\nonumber
\left(\begin{array}{cc}
\textbf{A}_{11} & \textbf{A}_{12}\\
\textbf{A}_{21} & \textbf{A}_{22}
\end{array}
\right)^{-1} =\left(\begin{array}{cc}
\textbf{A}_{11}^{-1} + \textbf{A}_{11}^{-1}\textbf{A}_{12}(\textbf{A}_{22} - \textbf{A}_{21}\textbf{A}_{11}^{-1}\textbf{A}_{12})^{-1}\textbf{A}_{21}\textbf{A}_{11}^{-1} & -\textbf{A}_{11}^{-1}\textbf{A}_{12}(\textbf{A}_{22} - \textbf{A}_{21}\textbf{A}_{11}^{-1}\textbf{A}_{12})^{-1}\\
-(\textbf{A}_{22} - \textbf{A}_{21}\textbf{A}_{11}^{-1}\textbf{A}_{12})^{-1}\textbf{A}_{21}\textbf{A}_{11}^{-1} & (\textbf{A}_{22} - \textbf{A}_{21}\textbf{A}_{11}^{-1}\textbf{A}_{12})^{-1}
\end{array}
\right),
\end{equation}
\noindent
and
\begin{equation}
\nonumber
\left(\begin{array}{cc}
\textbf{A}_{11} & \textbf{A}_{12}\\
\textbf{A}_{21} & \textbf{A}_{22}
\end{array}
\right)^{-1} =\left(\begin{array}{cc}
(\textbf{A}_{11} - \textbf{A}_{12}\textbf{A}_{22}^{-1}\textbf{A}_{21})^{-1}& -(\textbf{A}_{11} - \textbf{A}_{12}\textbf{A}_{22}^{-1}\textbf{A}_{21})^{-1}\textbf{A}_{12}\textbf{A}_{22}^{-1}\\
-\textbf{A}_{22}^{-1}\textbf{A}_{21}(\textbf{A}_{11} - \textbf{A}_{12}\textbf{A}_{22}^{-1}\textbf{A}_{21})^{-1} & \textbf{A}_{22}^{-1}+\textbf{A}_{22}^{-1}\textbf{A}_{21}(\textbf{A}_{11} - \textbf{A}_{12}\textbf{A}_{22}^{-1}\textbf{A}_{21})^{-1}\textbf{A}_{12}\textbf{A}_{22}^{-1}
\end{array}
\right),
\end{equation}
\noindent
for generic real-valued $M\times M$ matrix $\textbf{A}_{11}$, $M\times (p+r)$ matrix $\textbf{A}_{12}$, $(p+r)\times M$ matrix $\textbf{A}_{21}$, and $(p+r)\times (p+r)$ matrix $\textbf{A}_{22}$. Apply this second identity to,
\begin{equation}
\nonumber
(\textbf{H}^{\prime}\textbf{H}) = \left(\begin{array}{ccc}
2\textbf{I}_{n}& \textbf{X} & \textbf{G}\\
\textbf{X}^{\prime} & \textbf{X}^{\prime}\textbf{X}+\textbf{I}_{p} & \textbf{X}^{\prime}\textbf{G}\\
\textbf{G}^{\prime} &  \textbf{G}^{\prime}\textbf{X} & \textbf{G}^{\prime}\textbf{G}+\textbf{I}_{r}\\
\end{array}
\right).
\end{equation}
\noindent
to produce
\begin{equation}
\nonumber
(\textbf{H}^{\prime}\textbf{H})^{-1} =\left(\begin{array}{cc}
(\textbf{F} - \textbf{K}\textbf{L}^{-1}\textbf{K}^{\prime})^{-1}& -(\textbf{F} - \textbf{K}\textbf{L}^{-1}\textbf{K}^{\prime})^{-1}\textbf{K}\textbf{L}^{-1}\\
-\textbf{L}^{-1}\textbf{K}^{\prime}(\textbf{F} - \textbf{K}\textbf{L}^{-1}\textbf{K}^{\prime})^{-1} & \textbf{L}^{-1}+\textbf{L}^{-1}\textbf{K}^{\prime}(\textbf{F} - \textbf{K}\textbf{L}^{-1}\textbf{K}^{\prime})^{-1}\textbf{K}\textbf{L}^{-1}
\end{array}
\right) 
\end{equation}
Since, we have that $\textbf{H}^{\prime}\textbf{w} = (\textbf{R}^{\prime},\textbf{P}^{\prime})^{\prime}$, Equation (\ref{firstvecrep}) from the main text follows immediately. Apply the first block inverse identity to $\textbf{F}-\textbf{K}\textbf{L}^{-1}\textbf{K}^{\prime}$ to obtain $\textbf{F}_{11}$, $\textbf{F}_{12}$, $\textbf{F}_{21}$, and $\textbf{F}_{22}$. Apply the Sherman-Morrison Woodbury identity \citep{johan} to obtain $\textbf{F}_{1}^{-1}$.

\section*{Appendix C: Enforcing Sparse Discrepancy Parameters Instead of Marginalizing Discrepancy Parameters}\label{bypass}
Theorem~\ref{thm:4} shows that posterior inference on $\bm{\zeta}$ can be interpreted as a type of regression when using $f(\bm{\zeta}\vert \textbf{z})$ for inference, which marginalizes across $\textbf{q}$. Another choice is to enforce sparsity on the discrepancy parameter (i.e., setting $\textbf{q}$ equal to zero) before marginalizing it out instead of after, which we do in this article. That is, one might instead use the informative point mass prior of $f(\textbf{q} ) = I(\textbf{q} =  \bm{0}_{n,1})$ leading one to use $f(\bm{\zeta}\vert \textbf{z},\textbf{q} = \bm{0}_{n,1})$ for inference \citep{glm-mcculloch}.\\

\noindent
\textit{Result: Suppose $Z_{i}\vert Y_{i}$ is independently distributed according to either (\ref{EF2}) of the main text with $i = 1,\ldots, n$. Let $\textbf{y}$ be defined as in Theorem~\ref{thm:4} of the main text with the added assumption that $f(\textbf{q}) = I(\textbf{q} = \bm{0}_{n,1})$. Then $\bm{\zeta}\vert \textbf{z}$ is cGCM$(\bm{\alpha}_{M},\bm{\kappa}_{M},\bm{\mu}_{q}, \textbf{H}, \pi_{*}, \textbf{D};\bm{\psi}_{M})$, with $\bm{\mu}_{q} = \bm{0}_{2n,1}$, and  $\bm{\alpha}_{M}$, $\bm{\kappa}_{M}$, $\textbf{H}$, $\textbf{D}$, and $\bm{\psi}_{M}$ defined in Theorem~\ref{thm:4}.}\\

\noindent
\textit{Proof:} We have from Theorem~\ref{thm:4},
\begin{align}
\nonumber
&f(\bm{\zeta},\bm{\theta}\vert \textbf{z},\bm{\alpha}_{\beta},\bm{\alpha}_{\eta},\bm{\kappa}_{\beta},\bm{\kappa}_{\eta},\textbf{q} = \bm{0}_{n,1}) \\
\nonumber
&\propto\hspace{5pt}\frac{\pi_{*}(\bm{\theta})}{\mathrm{det}\left\lbrace \textbf{D}(\bm{\theta})\right\rbrace}\mathrm{exp}\left[\bm{\alpha}_{M}^{\prime}\textbf{D}(\bm{\theta})^{-1}\left(\textbf{H}\hspace{6pt} \textbf{Q}\right)  \left(\begin{matrix}
\bm{\zeta}\\ \bm{0}_{n,1}
\end{matrix} \right)-\bm{\kappa}_{M}^{\prime}\bm{\psi}_{M}\left\lbrace\textbf{D}(\bm{\theta})^{-1}\left(\textbf{H}\hspace{6pt}\textbf{Q}\right)\left(\begin{matrix}
\bm{\zeta}\\ \bm{0}_{n,1}
\end{matrix} \right)\right\rbrace\right]\\
\noindent
&=\hspace{5pt}\frac{\pi_{*}(\bm{\theta})}{\mathrm{det}\left\lbrace \textbf{D}(\bm{\theta})\right\rbrace}\mathrm{exp}\left[\bm{\alpha}_{M}^{\prime}\textbf{D}(\bm{\theta})^{-1}\textbf{H}\bm{\zeta} -\bm{\kappa}_{M}^{\prime}\bm{\psi}_{M}\left\lbrace\textbf{D}(\bm{\theta})^{-1}\textbf{H}\bm{\zeta}\right\rbrace\right],
\end{align}
and when integrating across $\bm{\theta}$ we obtain the cGCM in the statement above.\\

\noindent
The result above shows that $f(\bm{\zeta}\vert \textbf{z}, \textbf{q} = \bm{0}_{n,1})$ is a conditional GCM, however, it is currently unknown how to simulate from the conditional GCM in many settings.

\singlespacing
\bibliographystyle{jasa} 
\bibliography{myref33}

\end{document}